\documentclass[review,sageh,times,doublespace]{sagej}

\usepackage{amsmath}
\usepackage{moreverb,url,xcolor}
\usepackage[acronym]{glossaries}
\usepackage{tcolorbox}
\tcbuselibrary{listings, breakable}
\usepackage{etoolbox}
\usepackage{array}
\usepackage{algorithmic}
\usepackage{algorithm}
\usepackage{cleveref}

\newacronym[plural=CPUs, longplural=Central Processing Units]{cpu}{CPU}{Central Processing Unit}
\newacronym[plural=GPUs, longplural=Graphics Processing Units]{gpu}{GPU}{Graphics Processing Unit}
\newacronym[plural=ASICs, longplural=Application-Specific Integrated Circuit]{asic}{ASIC}{Application-Specific Integrated Circuit}
\newacronym{ram}{RAM}{Random Access Memory}
\newacronym{tlb}{TLB}{Translation Lookaside Buffer}
\newacronym[plural=HOTs, longplural=Hashed Octrees]{hot}{HOT}{Hashed Octree}
\newacronym[plural=HQTs, longplural=Hashed Quadtrees]{hqt}{HQT}{Hashed Quadtree}
\newacronym{orb}{ORB}{Orthognal Recursive Bisection}
\newacronym{dtt}{DTT}{Dual Tree Traversal}
\newacronym{numa}{NUMA}{Non-Uniform Memory Access}
\newacronym{ai}{AI}{Aritificial Intelligence}
\newacronym{dag}{DAG}{Directed Acyclic Graph}
\newacronym{abi}{ABI}{Application Binary Interface}
\newacronym[plural=flops, longplural=Floating Point Operations]{flop}{FLOP}{Floating Point Operation}
\newacronym{hpc}{HPC}{High Performance Computing}
\newacronym{bsp}{BSP}{Bulk Synchronous Parallel}
\newacronym{mpi}{MPI}{Message Passing Interface}
\newacronym{jit}{JIT}{Just In Time}
\newacronym{cuda}{CUDA}{Compute Unified Device Architecture}
\newacronym{rdma}{RDMA}{Remote Direct Memory Access}
\newacronym{clt}{CLT}{Central Limit Theorem}
\newacronym{aot}{AOT}{Ahead of Time}
\newacronym[plural=ISAs, longplural=Instruction Set Architectures]{isa}{ISA}{Instruction Set Architecture}
\newacronym{raii}{RAII}{Resource Allocation is Initialisation}
\newacronym[plural=CCXs, longplural=Core Complexes]{ccx}{CCX}{Core Complex}
\newacronym[plural=CCDs, longplural=Core Complex Dies]{ccd}{CCD}{Core Complex Die}

\newacronym{svd}{SVD}{Singular Value Decomposition}
\newacronym{rsvd}{rSVD}{Randomised Singular Value Decomposition}
\newacronym{cpqr}{CPQR}{Column Pivoted QR}
\newacronym{qr}{QR}{QR}
\newacronym{id}{ID}{Interpolative Decomposition}
\newacronym{aca}{ACA}{Adaptive Cross Approximation}
\newacronym{cur}{CUR}{Pseudo-Skeletal}
\newacronym{hodlr}{HODLR}{Hierarchical Off-Diagonal Low Rank}
\newacronym{hbs}{HBS/HSS}{Hierarchical Block Separable/Hierarchical Semi Separable}
\newacronym{blas_m2l}{BLAS M2L}{BLAS Based Multipole to Local (M2L) Field Translation}
\newacronym{fft_m2l}{FFT M2L}{FFT Based Multipole to Local (M2L) Field Translation}
\newacronym{h}{$\mathcal{H}$}{$\mathcal{H}$ Matrix}
\newacronym{htwo}{$\mathcal{H}^2$}{$\mathcal{H}^2$ Matrix}

\newacronym{fma}{FMA}{Fused Multiply and Add}
\newacronym{sisd}{SISD}{Single Instruction Single Data}
\newacronym{simd}{SIMD}{Single Instruction Multiple Data}
\newacronym{mimd}{MIMD}{Multiple Instruction Multiple Data}
\newacronym{simt}{SIMT}{Single Instruction Multiple Threads}

\newacronym{ilp}{ILP}{Instruction Level Parallelism}
\newacronym{tlp}{TLP}{Thread Level Parallelism}
\newacronym{dlp}{DLP}{Data Level Parallelism}
\newacronym{clp}{CLP}{Core Level Parallelism}

\newacronym{dram}{DRAM}{Dynamic Random Access Memory}
\newacronym{neon}{Neon}{Arm SIMD Extensions}
\newacronym{sse}{SSE}{Streaming SIMD Extensions}
\newacronym{avx}{AVX}{Advanced Vector Extensions/Gesher New Instructions}
\newacronym{avx2}{AVX2}{Advanced Vector Extensions/Haswell New Instructions}
\newacronym{avx512}{avx-512}{advanced vector extensions 512 bit}

\newacronym{blas}{BLAS}{Basic Linear Algebra Subprograms}
\newacronym{l2l}{L2L}{Local to Local}
\newacronym{l2p}{L2P}{Local to Particle}
\newacronym{m2l}{M2L}{Multipole to Local}
\newacronym{m2p}{M2L}{Multipole to Particle}
\newacronym{m2m}{M2M}{Multipole to Multipole}
\newacronym{p2p}{P2P}{Particle to Particle}
\newacronym{p2m}{P2M}{Particle to Multipole}

\newacronym[plural=kiFMMs, longplural=Kernel Independent Fast Multipole Methods]{kifmm}{kiFMM}{Kernel Independent Fast Multipole Method}
\newacronym[plural=bbFMMs, longplural=Black Box Fast Multipole Methods]{bbfmm}{bbFMM}{Black Box Fast Multipole Method}
\newacronym[plural=FMMs, longplural=Fast Multipole Methods]{fmm}{FMM}{Fast Multipole Method}
\newacronym[plural=PDEs]{pde}{PDE}{Partial Differential Equation}
\newacronym{mfs}{MFS}{Method of Fundamental Solutions}
\newacronym[plural=BIEs, longplural=Boundary Integral Equations]{bie}{BIE}{Boundary Integral Equation}
\newacronym[plural=BEMs, longplural=Boundary Element Methods]{bem}{BEM}{Boundary Element Method}
\newacronym[plural=FFTs, longplural=Fast Fourier Transforms]{fft}{FFT}{Fast Fourier Transform}
\newacronym[plural=DFTs, longplural=Discrete Fourier Transforms]{dft}{DFT}{Discrete Fourier Transform}
\newacronym{epcc}{EPCC}{Edinburgh Parallel Computing Centre}
\newacronym[plural=LETs, longplural=Locally Essential Trees]{let}{LET}{Locally Essential Tree}

\newcommand{\hpc}{\acrshort{hpc}\xspace}

\newcommand{\lettree}{\acrshort{let}\xspace}

\newcommand{\lettreefull}{\acrfull{let}\xspace}

\newcommand{\mpi}{\acrshort{mpi}\xspace}

\newcommand{\fmm}{\acrshort{fmm}\xspace}

\newcommand{\kifmm}{\acrshort{kifmm}\xspace}

\newcommand{\kifmmfull}{\acrfull{kifmm}\xspace}

\newcommand{\fmms}{\acrshortpl{fmm}\xspace}

\newcommand{\cpu}{\acrshort{cpu}\xspace}
\newcommand{\cpus}{\acrshortpl{cpu}\xspace}

\def\real{ \mathbb{R}}

\def\rthree{\real^3}

\def \xbf{\mathbf{x}}
\def \ybf{\mathbf{y}}

\def \ybfj{\mathbf{y}_j}
\def \xbfi{\mathbf{x}_i}

\usepackage{eucal}
\def\bigO#1{\mathcal{O}\left(#1\right)}
\newcommand{\mathbsf}[1]{\mathsf{\mathbf{#1}}}
\newcommand{\textitbf}[1]{\textit{\textbf{#1}}}

\usepackage{xspace}
\newtcolorbox[auto counter, number within=section]{mybox}[2][]{%
  colback=white!5!white, colframe=gray,
  fonttitle=\bfseries,
  title=Box~\thetcbcounter: #2, label={#1},
  breakable
}

\newenvironment{mytablebox}[1]{%
  
  \refstepcounter{table} 
  \begin{tcolorbox}[
    breakable,
    colback=white!5!white, colframe=gray,
    fonttitle=\bfseries,
    left=2pt, right=2pt, top=4pt, bottom=4pt,
    boxsep=4pt,
    before skip=6pt,
    after skip=6pt,
    title=Table~\thetable:~#1
  ]
  \raggedright
}{%
  \end{tcolorbox}
}

\newlength{\notationlabelwidth}
\newcommand{\notationsection}[1]{%
  \par\smallskip
  \centering\underline{\textbf{#1}}\par
  \smallskip\raggedright
}

\newcommand{\notationentry}[2]{%
  \par\noindent
  \hangindent=\notationlabelwidth
  \hangafter=1
  \makebox[\notationlabelwidth][l]{#1}#2\par
}

\crefname{mybox}{Box}{Boxes}
\Crefname{mybox}{Box}{Boxes}

\newcommand\BibTeX{{\rmfamily B\kern-.05em \textsc{i\kern-.025em b}\kern-.08em
T\kern-.1667em\lower.7ex\hbox{E}\kern-.125emX}}

\def\journalname{The International Journal of High Performance Computing Applications}

\def\doi{10.1177/ToBeAssigned}

\def\volumeyear{2025}

\begin{document}

\runninghead{Kailasa}
\title{A Simple Communication Scheme for Distributed Fast Multipole Methods}

\author{Srinath Kailasa\affilnum{1, 2}}

\affiliation{\affilnum{1}Graphcore, United Kingdom\\
\affilnum{2}Department of Engineering, University of Cambridge, United Kingdom\\
}

\corrauth{Srinath Kailasa,
Graphcore,
11-19 Wine Street,
Bristol, England
BS1~2PH, UK.}

\email{srik@graphcore.ai}

\begin{abstract}
We present a simple hierarchical communication scheme for distributed \acrfullpl{fmm} based on \mpi neighborhood collectives and uniform trees. The method targets the common case of extending an existing high-performance shared-memory uniform-tree \fmm implementation to distributed memory with minimal redesign while preserving any shared memory optimizations. Benchmarks on the ARCHER2 supercomputer demonstrate that our method can scale to very large problem sizes, we demonstrate weak-scaling up to 3.2e10 uniformly distributed points on 512 nodes of the machine in our largest runs. Our simplifications based on uniform trees result asymptotically optimal communication under practical constraints that we document. We record departure from this scaling for non-uniform points, however we still obtain practically useful runtimes due to the ability to retain our shared memory optimizations.
\end{abstract}


\maketitle
\section{Introduction}

Fast algorithms for computing long-range interactions between large numbers of particles arise in a wide range of scientific applications, including gravitational $N$-body simulations, electrostatics, molecular dynamics, and boundary integral methods for elliptic \acrshortpl{pde}. These problems typically involve computing sums of the form
\begin{equation}
  \label{eq:field}
  f(\xbfi) = \sum_{j=1}^N K(\xbfi, \ybfj) s_j, \quad i = 1, \ldots, M,
\end{equation}
where $\{\ybfj\}_{j=1}^N \subset \mathbb{R}^d$ are `source' points with associated densities $s_j \in \mathbb{F}$, and $\{\xbfi\}_{i=1}^M \subset \mathbb{R}^d$ are the `target' points at which we wish to evaluate the field, $\mathbb{F} \in \{\mathbb{C}, \mathbb{R}\}$ and $d$ is the spatial dimension. The kernel $K(\cdot,\cdot): \mathbb{R}^d \times \mathbb{R}^d \to \mathbb{F}$ describes the interaction between source and target points. The prototypical kernel of interest in examining $N$-body problems is the three-dimensional Laplace kernel, $K(\cdot, \cdot): \rthree \times \rthree \rightarrow \real$,
\begin{equation}
  \label{eq:laplace_kernel}
  K(\xbf, \ybf) = \frac{1}{\|\xbf - \ybf\|} \> \> (d = 3),
\end{equation}
which models gravitational and electrostatic potentials.

Direct evaluation of \eqref{eq:field} requires $\bigO{NM}$ operations, making it computationally prohibitive at large scales. Moreover, \eqref{eq:field} implies a global data dependency between all source and target points. However, when the kernel admits a low-rank structure for well-separated point sets, hierarchical algorithms such as the \acrfull{fmm}~\citep{greengard1987fast} or the closely related $\mathcal{H}$-matrices~\citep{hackbusch1989fast,hackbusch1999sparse} can reduce the computational cost to $\bigO{N+M}$ in the best case.

These methods work by recursively partitioning the computational domain using hierarchical data structures (commonly octrees for $d=3$ or quadtrees for $d=2$), and splitting the sum in \eqref{eq:field} into near-field and far-field contributions for each box in the hierarchy containing target particles:
\begin{equation}
  f(\xbfi) = \sum_{\ybfj \in \text{Near}(\xbfi)} K(\xbfi, \ybfj) s_j + \sum_{\ybfj \in \text{Far}(\xbfi)} K(\xbfi, \ybfj) s_j.
  \label{eq:sec:introduction:near_far_split}
\end{equation}
The near-field interactions are evaluated directly, while the far-field interactions are approximated using compressed representations.

Despite reducing algorithmic complexity, hierarchical methods do not eliminate the global data dependencies of the original $N$-body problem. In distributed memory, a naive implementation typically results in expensive all-to-all communication, with worst-case cost $\bigO{P^2}$, where $P$ is the number of processes. Communication complexity therefore limits parallel scalability and must be considered when adapting $N$-body software to current and emerging exascale hardware.

A landmark result by Lashuk et al.~\citep{lashuk2009massively} demonstrated an approach that reduced the runtime communication complexity for \fmms to $\bigO{\sqrt{P}}$, where $P$ is the number of processes. More recently, Ibeid et al.~\citep{ibeid2016performance} showed that communication can in principle scale as low as $\bigO{\log{P}}$ when the communication pattern mirrors the hierarchical structure of the algorithm. However, for practical distributed implementations, asymptotic communication complexity is only one part of the story. Communication schemes must also be simple to implement, portable across \mpi libraries and machine topologies, and compatible with the shared-memory optimizations critical for single-node performance. This is particularly relevant on modern \hpc systems, where hardware is deeply hierarchical \citep{kim2008technology} and overall performance is strongly controlled by the computational intensity of the underlying shared-memory kernels \citep{williams2009roofline}. In this setting, a method that relies only on common collective operations and a precomputed communication layout while preserving efficient local kernels can be more valuable than one that pursues the strongest asymptotic bound at the cost of substantial implementation complexity.

In this paper, we contribute:

\begin{enumerate}
  \item A communication strategy for distributed \fmm using only standard MPI-4 collective operations and a precomputed communication layout, with neighborhood collectives for local exchange, designed to preserve shared-memory optimizations developed for single-node \fmms and to align with existing software using uniform trees and common spatial encoding schemes.
  \item A communication analysis of the distributed \fmm phase of our implementation showing a favorable bound of $\bigO{\log P + (N/P)^{2/3}}$ under explicit practical assumptions, and identifying the conditions under which the logarithmic global term is expected to be relevant.
  \item Large-scale benchmarks on ARCHER2 studying weak scaling for both uniform and non-uniform distributions, together with a practical strong-scaling study. We find our method capable of practically computing large problem sizes. Despite not adhering to asymptotic optimality in communication in all cases, we are able to compute the \fmm over large numbers (3.2e10) of uniform/non-uniform points in practically useful times (of the order of 10s of seconds).
\end{enumerate}

We provide an overview of the \fmm in distributed memory in Section~\ref{sec:fmm}, with a high-level overview of the sequential algorithm in Section \ref{sec:fmm:sub:overview}, the importance of shared memory efficiency for distributed implementations in Section \ref{sec:fmm:sub:shared}, and a review of efforts to extend the algorithm to distributed memory in Section \ref{sec:fmm:sub:distributed} to build the required context of our own contributions. Section~\ref{sec:algorithm} describes our approach, Section~\ref{sec:analysis} presents a theoretical complexity analysis, and Section~\ref{sec:benchmarks} reports our performance results on the \cpu-based ARCHER2 system hosted at the \acrfull{epcc}. In Section~\ref{sec:discussion}, we contextualize these results relative to existing work, and conclude with future directions in Section~\ref{sec:conclusion}.

\section{Fast Multipole Methods}\label{sec:fmm}


\subsection{High Level Overview of the \fmm}\label{sec:fmm:sub:overview}

The \fmm has come to embody a wide variety of closely related methods, which though sharing a common algorithmic structure, differ in their approach to field approximation and compression of the far-field in \eqref{eq:sec:introduction:near_far_split}. Additionally, implementation details such as the use of hardware accelerators, parallel runtimes, and programming languages lead to a further diversity in variant \fmms. To avoid this complexity, we describe the structure common to all techniques, and defer to the literature and sources therein for details of both the algorithmic structure \citep{ying2004kernel,greengard1987fast,fong2009black, darve2000fast, anderson1992implementation, martinsson2007accelerated} in addition to approaches for the optimization of the implementation, which often depend closely on the chosen algorithmic approach \citep{kailasa2024m2l,malhotra2015pvfmm,lashuk2009massively,wang2021exafmm,yokota2012tuned,ltaief2014data,abduljabbar2014asynchronous,gumerov2008fast, chandramowlishwaran2010optimizing,blanchard2015scalfmm,messner2012optimized,bramas2020tbfmm}.

Most modern implementations of \fmms are built using hierarchical data structures called \acrfullpl{hot} for problems in three spatial dimensions (or \acrfullpl{hqt} in two dimensions), pioneered by Warren and Salmon \citep{warren1993parallel}. This includes the most recent \texttt{exafmm} variant \citep{wang2021exafmm}, \texttt{pvfmm} \citep{malhotra2015pvfmm}, and our own \texttt{kifmm-rs} \cite{kailasa2025joss}. Alternative approaches based on graph bisection \citep{abduljabbar2014asynchronous} and methods specialized for task-based runtimes \citep{blanchard2015scalfmm} also exist. We do not discuss their communication schemes in detail because they are closely tied to the underlying data structures and runtimes used in those cases.

For ease of reference, we employ the notation of Lashuk et al. \citep{lashuk2009massively}, summarized in Table \ref{tab:notation}. Consider a problem in $\rthree$. We enclose all source and target points in a cube of sufficient size, then construct an octree $T$ by refining the cube into eight octants, or \textit{boxes}, which are recursively subdivided until a user-defined \textit{depth}, $d$.

This corresponds to an \fmm with \textit{uniform refinement}, in which each leaf box is the same size. We associate a \textit{level}, $l \in [0, d]$, with each box $\beta$ in $T$, where $l = d$ is the deepest level and $l = 0$ corresponds to the \textit{root level} occupied by the bounding cube. \textit{Adaptive refinement} is also possible, in which adjacent leaf boxes (that is, boxes that share a vertex, edge, or face) may differ in size. In that case boxes are refined until no leaf box contains more than $q$ sources/targets. We denote the set of leaf boxes by $L \subset T$.

\begin{figure}
  \centering
  \includegraphics[width=0.7\textwidth]{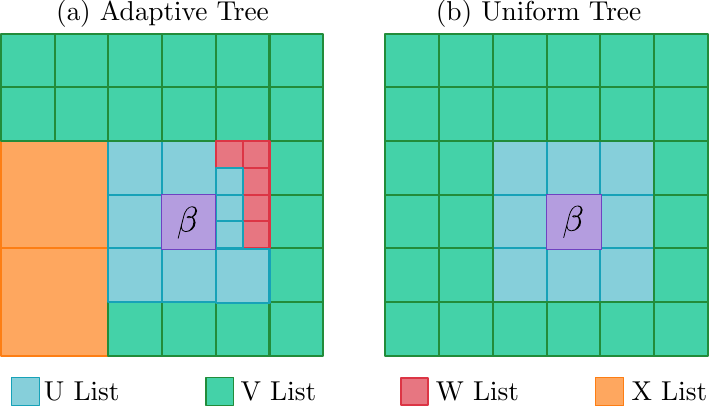}
  \caption{Interaction lists for a box $\beta \in T$, where $T$ is a quadtree for a problem in $\mathbb{R}^2$, shown for clarity. Panel (a) shows an adaptively refined tree and panel (b) a uniformly refined tree.}
  \label{fig:interaction_lists}
\end{figure}

We associate with each $\beta \in T$ four so-called \textit{interaction lists}. These are lists of boxes whose interactions with each $\beta$ are coupled during the course of the \fmm algorithm, called the U, V, X and W lists in the terminology of \citep{ying2004kernel}, which we illustrate in Figure \ref{fig:interaction_lists}.

In the uniform case, only the U and V lists are defined. We begin by defining the colleagues, $\mathcal{C}(\beta)$, of $\beta$ as boxes which are adjacent to $\beta$ and at the same level. If $\beta$ is a \textit{leaf box} such that it is not recursively subdivided into more boxes, its U list consists of its colleagues, $\mathcal{C}(\beta)$, in addition to $\beta$ itself. For all boxes $\beta \in T$, the V list consists of boxes at the same level whose parents are colleagues of $\beta$'s parent.

In the adaptive case, the U list of a leaf box, $\beta$, now consists of all adjacent boxes, not just colleagues, since adjacent boxes may not be at the same level. The definition of the V list remains the same. In addition we define W and X lists for $\beta$. The W list consists of boxes that are descendants of colleagues of $\beta$ and are themselves not adjacent to $\beta$. The X list consists of boxes for which $\beta$ appears in their own W list.

\begin{figure}
  \centering
  \includegraphics[width=0.7\textwidth]{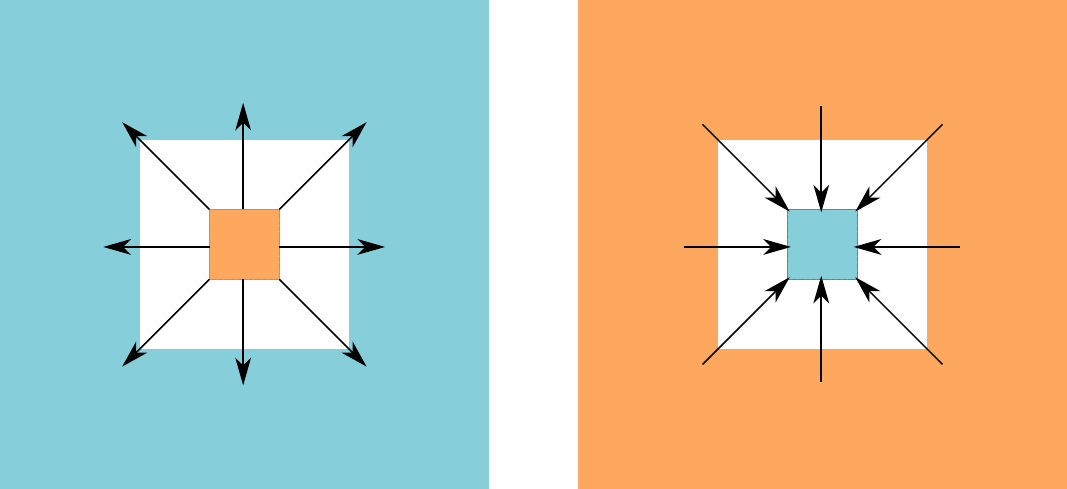}
  \caption{We illustrate a problem in $\real^2$. The left plot shows the well-separated blue region in the exterior of an orange source box where the representation $\mathbsf{u}$ for the source box is valid. The right plot shows the blue region of validity in the interior of a target box where $\mathbsf{d}$ summarizes the field from well-separated source boxes in the orange exterior. The arrows represent the outgoing and incoming nature of the representations $\mathbsf{u}$ and $\mathbsf{d}$ respectively.}
  \label{fig:multipole_and_local}
\end{figure}

Each box $\beta \in T$ is associated with two vectors, $\mathbsf{u}$ and $\mathbsf{d}$. With reference to Figure \ref{fig:multipole_and_local}, the vector $\mathbsf{u}$ is understood as a compressed representation of the field due to densities contained \textit{within the box}, which is accurate outside the region covered by $\mathcal{C}(\beta)$. The vector $\mathbsf{d}$ is understood as a compressed representation of the field due to densities \textit{located outside} $\mathcal{C}(\beta)$, and therefore describes the field evaluated at target points contained within $\beta$ due to source densities in its far-field.

Additionally, leaf boxes, $\beta \in L$, are also associated with vectors $\mathbsf{x}$ -- the target points contained in $\beta$, $\mathbsf{y}$ -- the source points contained in $\beta$, $\mathbsf{s}$ -- the associated source densities, and $\mathbsf{f}$ -- the field evaluated at target points in $\beta$.

When using \acrshortpl{hot} or \acrshortpl{hqt}, boxes are encoded using a scheme that preserves spatial locality, most commonly Morton or Hilbert keys. These are typically stored as 64-bit integers that represent the relative positions of boxes within the domain, with three bits encoding the box's location at each level of refinement. A direct mapping of these keys to memory addresses performs poorly for adaptive point distributions, since many boxes are empty. Therefore, keys corresponding to non-empty boxes are \textit{hashed} so that adjacent hashes correspond to adjacent memory addresses. We use Morton keys in our implementation \citep{sundar2008bottom}; when sorted, they correspond to a pre-order traversal of the tree and therefore provide an efficient way to index data associated with each box during the recursive algorithm, while also ensuring that spatially local data corresponds to geometrically local boxes.

The reduced computational complexity of the \fmm is due to the interaction lists and recursive tree structure, which enable the field, $\mathbsf{f}$, to be evaluated at each leaf box, $\beta \in L$, by coupling interactions between $\beta$ and only a \textit{bounded} number of other boxes. Linear complexity is obtained as we refine the hierarchical data structure such that there are $\bigO{N}$ leaf boxes.

The sequential algorithm, designed for shared memory, is summarized for the uniform and adaptive cases in Algorithm \ref{alg:fmm}. It relies on a number of linear operators, described in Table \ref{tab:notation}, which translate between the representations of the field due to $\mathbsf{s}$, described by $\mathbsf{u}$ and $\mathbsf{d}$, and the final evaluated field $\mathbsf{f}$. The structure and construction of these operators is the main point of difference between various \fmms, with corresponding trade-offs in implementation, performance, and accuracy; some of these trade-offs are summarized in \citep{yokota2013fmm}. Construction methods range from algebraic approaches, based on the construction and compression of blocks of a matrix corresponding to \eqref{eq:field}, to purely analytical techniques that employ potential theory to derive kernel-specific formulae. Our software \cite{kailasa2025joss}, which we use for the benchmarking reported in this paper is based on the algebraic \kifmmfull \citep{ying2004kernel}.

\begin{mytablebox}{Summary of notation.}
\footnotesize
\notationsection{Box Data}
\notationentry{$\beta, \alpha$}{Unique representation of a box}
\notationentry{$\mathbsf{x}$}{Target points (leaf boxes only)}
\notationentry{$\mathbsf{y}$}{Source points (leaf boxes only)}
\notationentry{$\mathbsf{s}$}{Source densities at source points (leaf boxes only)}
\notationentry{$\mathbsf{f}$}{Potential at target points (leaf boxes only)}
\notationentry{$\mathbsf{u}$}{Multipole representation}
\notationentry{$\mathbsf{d}$}{Local representation}

\notationsection{Operators}
\notationentry{$S_\beta$}{Source-to-Up (S2U) or Particle-to-Multipole (P2M) operator}
\notationentry{$U_{\mathcal{P}(\beta), \beta}$}{Up-to-Up (U2U) or Multipole-to-Multipole (M2M) operator}
\notationentry{$T_{\beta, \alpha}$}{Up-to-Down (U2D) or Multipole-to-Local (M2L) operator}
\notationentry{$D_{\beta, \mathcal{P}(\beta)}$}{Down-to-Down (D2D) or Local-to-Local (L2L) operator}
\notationentry{$Q_{\beta, \alpha}$}{Source-to-Down (S2D) or Particle-to-Local (P2L) operator}
\notationentry{$R_{\beta, \alpha}$}{Up-to-Target (U2T) or Multipole-to-Particle (M2P) operator}
\notationentry{$E_{\beta}$}{Down-To-Target (D2T) or Local-to-Particle (L2P) operator}
\notationentry{$K_{\beta, \alpha}$}{Source-To-Target (S2T) or Particle-to-Particle (P2P) operator}

\notationsection{Tree Data}
\notationentry{$T$}{Hierarchical Tree}
\notationentry{$L$}{Leaf boxes of $T$ where $L \subset T$; boxes $\beta \in L$ are defined by not being further subdivided}
\notationentry{$\mathcal{A}(\beta)$}{All ancestor boxes of $\beta \in T$}
\notationentry{$\mathcal{L}(\beta)$}{Tree level of a box $\beta \in T$}
\notationentry{$\mathcal{C}(\beta)$}{Colleagues of a box $\beta \in T$}
\notationentry{$\mathcal{P}(\beta)$}{Parent of a box $\beta \in T$}
\notationentry{$\mathcal{K}(\beta)$}{Children of a box $\beta \in T$}
\notationentry{$\mathcal{N}(\beta)$}{The near field of a box $\beta \in T$, defined by a halo of boxes $\alpha$ such that they are contained in $\mathcal{C}(\beta)$ where $\mathcal{L}(\alpha) = \mathcal{L}(\beta)$}
\notationentry{$\mathcal{F}(\beta)$}{The far field of a box $\beta \in T$, defined as a complement of $T$ with respect to $\mathcal{N}(\beta)$}
\notationentry{$\mathcal{U}(\beta)$}{Boxes adjacent to box $\beta$, including $\beta$ itself}
\notationentry{$\mathcal{V}(\beta)$}{Boxes $\alpha \in T$ such that $\alpha \in \mathcal{K}(\mathcal{C}(\mathcal{P}(\beta)))$, $\mathcal{L}(\alpha) = \mathcal{L}(\beta)$, and $\alpha$ and $\beta$ are not adjacent}
\notationentry{$\mathcal{W}(\beta)$}{Boxes $\alpha \in T$ such that $\alpha \in \mathcal{N}(\beta)$ and the $\mathbsf{u}$ of $\alpha$ converges within $\beta$}
\notationentry{$\mathcal{X}(\beta)$}{Boxes $\alpha \in T$ such that $\beta \in \mathcal{N}(\alpha)$, and the densities $\mathbsf{s}$ are evaluated to construct $\mathbsf{d}$ for $\beta$}

\vspace{0.5em}
\footnotesize\noindent\emph{Adapted from Table 1} \citep{lashuk2009massively}.
\vspace{0.5em}
\label{tab:notation}
\end{mytablebox}

\begin{algorithm}

  \begin{algorithmic}
  \STATE \textbf{Input}: $\{\mathbsf{x}_i, \mathbsf{y}_i, \mathbsf{s}_i\}_{i=1}^{N_{\text{oct}}}$ - An octree with $|T| = N_{\text{oct}}$ octants.
  \STATE \textbf{Output}: $\{\mathbsf{f}_i\}_{i=1}^{N_{\text{leaf}}}$ - the potentials evaluated at target points at each of $|L| = N_{\text{leaf}}$ leaf octants.

  \STATE // \textbf{(1)} The \textitbf{upward pass}, traverses the tree, $T$, in post-order - from finest to coarsest boxes, level by level

  \STATE $\forall \beta \in L$ : $\mathbsf{u}_\beta = S_\beta \mathbsf{s}_\beta$ // \textbf{(1a) S2U}:source-to-up step
  \STATE $\forall \beta \in T$: $\mathbsf{u}_{P(\beta)} += U_{P(\beta), \beta} \mathbsf{u}_{\beta}$ // \textbf{(1b) U2U}: up-to-up step

  \STATE
  \STATE // \textbf{(2)} The \textitbf{downward pass}, traverses the tree, $T$, in pre-order - from the coarsest to the finest boxes, level by level

  \STATE $\forall \beta \in T$: $\forall \alpha \in \mathcal{V}(\beta)$: $\mathbsf{d}_\beta += T_{\beta, \alpha} \mathbsf{u}_\beta$ // \textbf{(2a) VLI}: V-list interactions

  \IF{$T$ is adaptive}
      \STATE $\forall \beta \in T$: $\forall \alpha \in \mathcal{X}(\beta)$: $\mathbsf{d}_\beta += Q_{\beta, \alpha} \mathbsf{s}_\alpha$ // \textbf{(2b) XLI}: X-list interactions
  \ENDIF

  \STATE $\forall \beta \in T$: $\mathbsf{d}_{\beta} += D_{\beta, P(\beta)} \mathbsf{d}_{P(\beta)}$ // \textbf{(2c) D2D}: down-to-down step

  \STATE
  \STATE // \textbf{(3)} The \textitbf{leaf level} boxes are finally handled, where the far-field is accumulated, and the near field directly summed.

  \IF{$T$ is adaptive}
      \STATE $\forall \beta \in L$: $\forall \alpha \in \mathcal{W}(\beta)$: $\mathbsf{f}_\beta += R_{\beta, \alpha} \mathbsf{u}_\alpha$ // \textbf{(3a) WLI}: W-list interactions
  \ENDIF

  \STATE $\forall \beta \in L$: $f_\beta += E_\beta d_\beta$ // \textbf{(3b) D2T}: down-to-target step

  \STATE $\forall \beta \in L$: $\forall \alpha \in U(\beta)$: $\mathbsf{f}_\beta += K_{\beta, \alpha} \mathbsf{s}_\alpha$ // \textbf{(3c) ULI}: U-list interactions (direct kernel evaluations)

  \end{algorithmic}
  \caption{ \textbf{(Shared Memory FMM)}: We adapt the compact specification of Algorithm 1 in \citep{lashuk2009massively}, but highlight steps unique to the adaptive \fmm.}
  \label{alg:fmm}
\end{algorithm}

\subsection{Shared Memory Optimizations and Their Role in Distributed \fmms}\label{sec:fmm:sub:shared}

In this section, we review key memory and parallelism considerations, which we aim to preserve when scaling to distributed systems.

Examining Algorithm \ref{alg:fmm}, firstly in terms of coarse-grained parallelism, we note that the (exact) \textbf{ULI} step is independent from the remainder of the upward/downward passes for approximate field calculations, and can be performed asynchronously. Furthermore the \textbf{VLI} and \textbf{XLI} steps are independent level-wise, after the upward pass. Within each \textbf{VLI} interaction for boxes $\beta \in T$ at a given level $l$ during the downward pass, many interactions across source and target boxes share translational symmetry and share a common Up-to-Down (U2D) operator during the \textbf{VLI} step, this is also true for \textbf{U2D} operations \textit{within} the V-list of a single target box, and techniques that exploit this symmetry are at the core of \citep{kailasa2024m2l,messner2012optimized,gumerov2008fast,malhotra2015pvfmm} – as they enable optimizations designed to reduce cache misses through the sharing of data amongst multiple \textbf{U2D} calculations. As already mentioned, fine grained parallelism is exposed for the \textbf{ULI} step, which is well suited for modern \acrshort{simd} or \acrshort{simt} machines. Coalescing of the \textbf{U2U} and \textbf{D2D} operations is also possible across multiple sets of sibling octants (those which share a common parent), however is less significant for runtime as memory accesses are spatially local, corresponding to the geometric locality of the corresponding boxes involved in these operations.

With reference to Figure \ref{fig:interaction_lists} coupling interactions between a box $\beta$ and boxes in its various interaction list naturally leads to cache-inefficient memory accesses over non-contiguously stored data in a \acrshort{hot} approach. The efficient handling of these memory accesses is often the greatest barrier to achieving high computational intensity in the shared memory portion of an \fmm implementation. The memory hierarchies of modern processors impose significant costs on memory accesses that result from cache misses, and often determine the end-to-end performance of an implementation. As a result the \textbf{VLI}, \textbf{XLI} and \textbf{WLI} steps in Algorithm \ref{alg:fmm} require careful design to ensure that implementations are not limited by the slowness of inefficient memory access \citep{messner2012optimized,kailasa2024m2l,malhotra2015pvfmm,takahashi2012optimizing,gumerov2008fast,fong2009black}.

The \fmm was originally introduced to avoid direct kernel evaluations using \eqref{eq:field}, and minimize the \textbf{ULI} step, however this step is highly suited to the structure of modern \acrshort{simd} and \acrshort{simt} processors – due to the high rate of data re-use available for each target particle in \eqref{eq:field}. Therefore, the performance of implementations is controlled by the depth of the associated tree. A deeper tree corresponding to a smaller number of $q$ particles for $\beta \in L$, resulting in more \textbf{VLI} and \textbf{XLI} steps, which are bandwidth bound, against shallower trees which result in larger \textbf{ULI} steps which we can write with high operational intensity and are compute bound. Indeed, these steps, in particular the \textbf{VLI} and \textbf{ULI} steps, dominate the total runtime of an \fmm, often exceeding 90 \% of wall time \citep{chandramowlishwaran2010optimizing,kailasa2024m2l}.

In the uniform case, one can take advantage of stencil based methods to re-form the \textbf{VLI} step such that spatial and temporal locality is maintained in memory access, and have been explored in \citep{kailasa2024m2l,gumerov2008fast,messner2012optimized,malhotra2015pvfmm,takahashi2012optimizing}. This results from the fact that the U and V are simply halos around each box and known a priori, and simply checked for existence at runtime (i.e. whether or not an expected box in U, or V, contains any particle data).

This allows for significant memory optimizations, such as the pre-allocation of contiguous buffers to store box data required during the \textbf{ULI} and \textbf{VLI} steps and index maps to rapidly look up data associated with interactions lists for a given target box. However, in the adaptive case interaction lists depend on the particular particle distribution – and must be calculated at runtime. Because of this, we favor shallow uniformly refined trees that minimize memory-bound interactions and delegate irregularity to compute-bound kernel evaluations, where \acrshort{simd} acceleration is highly effective. We justify this approach for the distributed memory case by experiment in Section \ref{sec:benchmarks}.

\subsection{Distributed Memory Parallelism}\label{sec:fmm:sub:distributed}

We review the communication requirements of the distributed \fmm when \acrshortpl{hot}/\acrshortpl{hqt} are used as the underlying hierarchical data structures. We describe the algorithm in three spatial dimensions using \acrshortpl{hot}. The input to a distributed \fmm is a set of $N$ source points and $N$ target points, which without loss of generality we take to be the same set, together with $N$ densities associated with the source points. These points are input in arbitrary order, so each process $i \in [0,P-1]$ initially stores approximately $N/P$ points, where $P$ is the total number of processes used. By `process' we mean a coarse-grained unit of computation with its own memory space and execution state, such as an \mpi process. The output is a distributed vector $\mathbsf{f}$, where the portion $\mathbsf{f}_i$ on each process contains the evaluated field for the target points stored on process $i$.

\subsubsection{Tree Construction and Domain Partitioning}

\begin{figure}
  \includegraphics[width=\textwidth]{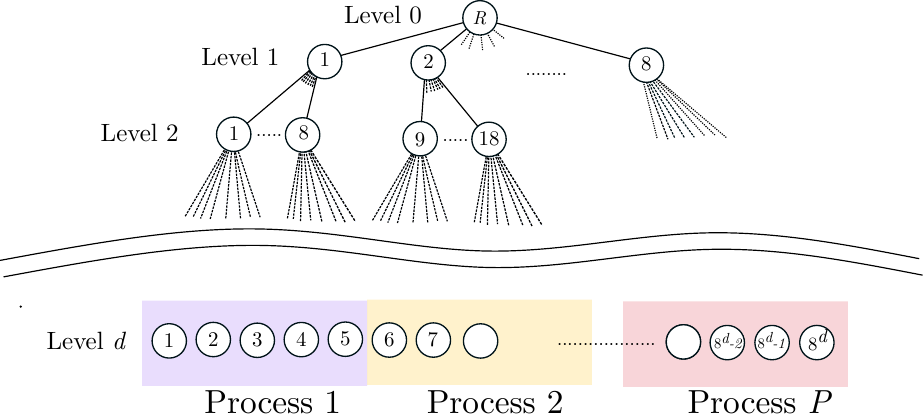}
  \caption{We illustrate the partition of a computational domain $\Omega$ discretized with a uniform linear \acrfull{hot} discretized to a depth $d$, such that each process $i \in [0,P-1]$ is responsible for a subdomain $\Omega_i$ defined by the leaf boxes such that $\Omega$ is formed from their disjoint union $\Omega = \bigcup_{i = 1}^P \Omega_i$. We illustrate that a spatial encoding scheme that preserves geometric locality amongst boxes leads to a partition which reflects this spatial locality in the logical proximity of neighboring processes.}
  \label{fig:let}
\end{figure}

When using \acrshortpl{hot} we follow the approach for the construction of distributed \textit{linear} octrees first described in \citep{sundar2008bottom}. Here, the computational domain $\Omega$ is partitioned across processes such that $\Omega = \bigcup_{i=1}^P \Omega_i$, where each subdomain $\Omega_i$ is defined by the volume covered by the \textit{leaf boxes} assigned to it. We illustrate such a tree in Figure \ref{fig:let}. We define the \lettreefull for each process as,

\begin{equation}
  \label{eq:let}
  \text{LET}(i) = \bigcup_{\beta \in [ L_i \bigcup \mathcal{A}(L_i)]} \mathcal{I}(\beta).
\end{equation}

This is the subtree of the global \acrshort{hot} containing the boxes it requires to compute its local contribution to the field, $\mathbsf{f}_i$, at each process. It contains a union of data associated with the combined interaction lists, $\mathcal{I}(\beta)$, of all the leaves, $L_i$, local to that process in addition to all of their collective ancestors $\mathcal{A}(L_i)$. The \lettree is seen to be the minimum set of box data required to evaluate $\mathbsf{f}_i$.

\subsubsection{Communication Phases in Distributed FMM}

When using \acrshortpl{hot}, the leaf boxes are found by performing a spatial encoding; here we describe a scheme based on Morton keys \citep{sundar2008bottom}. A similar approach would apply to alternative spatial encoding schemes. We assign each point a Morton key encoded at the maximum level, or depth $d$. This is pleasingly parallel over all processes, and as such requires $O(N)$ work. Once these keys have been found, there are five major communication-intensive steps, which can be distinguished by whether the communication is executed during setup or during runtime of the \fmm algorithm:

\begin{mybox}[box:comm-steps]{Distributed Communication Steps}
\begin{enumerate}
  \item \textbf{(Setup)}: A distributed sort of the input points based on their spatial encoding, such that geometrically close points are assigned to logically close processes.
  \item \textbf{(Setup)}: The construction of \acrshortpl{let} on each process to construct the interaction lists for boxes under their control.
  \item \textbf{(Setup)}: The communication of densities, $\mathbsf{s}$, required during the \textbf{ULI} step (in the uniform case), as well as the \textbf{XLI} step (in the adaptive case). This will necessarily be between logically close processes due to Step (1).
  \item \textbf{(Setup)}: Load balancing input to balance workload among partitions.
  \item \textbf{(Runtime)}: The communication of the $\mathbsf{u}$ vectors calculated during the upward pass to processes that require this data during the \textbf{VLI} steps (in the uniform case), as well as the \textbf{WLI} steps (in the adaptive case).
\end{enumerate}
\end{mybox}

\subsubsection{Distributed Sorting Strategies}

A simple approach for Step (1) in Box \ref{box:comm-steps} is provided by \textitbf{Samplesort} \citep{blelloch1991comparison}. Here, we identify $P$ buckets associated with each process such that all keys in the $i$'th bucket are less than or equal to all keys in the $(i+1)$'th bucket for $i \in [1, P]$. We distribute the leaves $L$ into appropriate buckets and sort the local buckets, $L_i$, to end up with a globally sorted $L$. In our software we use a randomized scheme for selecting $P-1$ splitters to define these buckets. Here, the Morton keys of $b$ points are selected from each of $i \in [1, P]$ processes and gathered at a nominated process. Once a local sort has been applied to this subset of entries, one can select $P-1$ entries by choosing every $b$'th entry as a splitter. The splitters can be used to form an all-to-all communication between all $P$ processes to redistribute local portions of point data into their appropriate bucket.

Using modern implementations of \mpi the dominant costs of the sort have a complexity of,

\begin{equation}
    \bigO{t_r (\frac{N}{P} \log{\frac{N}{P}}) + t_i b \log{P} + t_i \bar{m} P },
\end{equation}

where the first term describes the cost of the local sorts at each node, each initialized with $\bigO{N/P}$ keys, with $t_r$ describing the slowness of local memory accesses as $t_r = 1/\text{bandwidth}$. The second term describes the cost of gathering samples of size $b$ from each process using schemes based on binomial trees or recursive doubling, where $t_i$ describes the cost of communication over the interconnect as $t_i = 1/\text{bandwidth}$. The final term describes the cost of the all-to-all operation required to distribute entries into their respective buckets, where $\bar{m}$ is the mean message size. This corresponds to an \texttt{MPI\_Alltoallv} operation, and for large message sizes can be accomplished in $\bigO{P}$ communication steps in modern \mpi implementations \citep{kang2020improving}. As such, this forms the most expensive part of our implementation. However, as our focus is on the runtime phases of the \fmm in this paper we treat this cost as a fixed startup cost that is paid only once when setting up the tree. We acknowledge that asymptotically better algorithms exist, see for example \citep{sundar2013hyksort}.

\subsubsection{Load Balancing}

Our implementation targets uniform and near uniform point distributions, such that we can uniform trees. In which case The \acrfull{clt} implies that the sample quantiles used as splitters the parallel sort step increasingly approximate the true quantiles as the sample size \( b \) grows. Specifically, the standard deviation of the splitter positions decreases proportionally to \(1/\sqrt{b}\). This ensures that the buckets formed by these splitters contain approximately equal numbers of entries, with deviations from the ideal size \( N/P \) diminishing with larger sample sizes.

In the context of the uniform \fmm the estimated buckets correspond to partitions of the points such that each bucket contains an approximately equal number of points. For uniform or approximately uniform point distributions, such as points distributed evenly throughout the problem domain or uniformly on a surface without sharp features, the proposed random sampling approach therefore leads to an approximately load balanced partition of points across \mpi processes that improves with the number of samples. The uniformity of the distribution means that this also corresponds to an approximately load balanced distribution of leaf boxes $L_i \subset L$ across processes $i \in [0,P-1]$

This analysis does not to non-uniform point distributions, and in this case we expect a significant load imbalance. In our experiments we demonstrate that practically useful large scale computations are still achievable in practice despite this – without the need for further specialized load-balancing algorithms.

\subsubsection{Runtime Communication Scheme of Lashuk et al}

Lashuk et al. \citep{lashuk2009massively} split runtime communication in Step (5) in Box \ref{box:comm-steps} into two parts, in the first part they perform the upward pass independently over each \lettree, acknowledging that for boxes close to the root of the tree their $\mathbsf{u}$ vectors will only contain partial results, which must be synchronized across \textit{all} processes. In order to avoid a global all-to-all, they use a customized `reduce-scatter' scheme based on point-to-point communication, discussed in Algorithm 3 of \citep{lashuk2009massively}, using a classical hypercube communication scheme. This algorithm has a communication complexity of,

\begin{equation}
  \label{eq:lashuk_runtime_complexity}
  \bigO{t_l \log{P} + t_i \bar{m} (3 \sqrt{P} - 2)},
\end{equation}

where $t_l$ is a constant associated with the latency of establishing communication between two processes over $\log{P}$ communication rounds, and $t_i$ is constant related to the bandwidth constraints of a \hpc system of sending a mean of $\bar{m}$ messages between processes during each round. The bandwidth limited nature of modern \hpc systems leads to the dominant cost being proportional to $\sqrt{P}$. In the uniform case $\bar{m}$ can be estimated as $\left(\frac{N}{P}\right)^{2/3}$, with an argument relating the surface area to volume ratio of the subdomain controlled by each process, $\Omega_i$, which is proportional to the octant data required by other processes. This leads to an estimate of the dominant cost in communication complexity being,

\begin{equation}
  \label{eq:lashuk_runtime_complexity_uniform}
  \bigO{\left(\frac{N}{P}\right)^{2/3} \sqrt{P}}
\end{equation}

Once completed the downward passes can be performed independently on each process to calculate $\mathbsf{f}_i$ for their local portion of target points. This scheme maintains any shared memory optimizations developed for both the upward and downward passes, however introduces redundant computations for $\mathbsf{d}$ at higher levels of the tree during the \textbf{VLI} step – which are repeated across all processes which share many common ancestor boxes close to the root. Data dependencies are explicitly resolved at runtime, including the discovery of coupled interactions and the associated allocations required to store interaction list box data.

\subsubsection{Runtime Communication Scheme of Ibeid et al}

\begin{figure}
  \includegraphics[width=\textwidth]{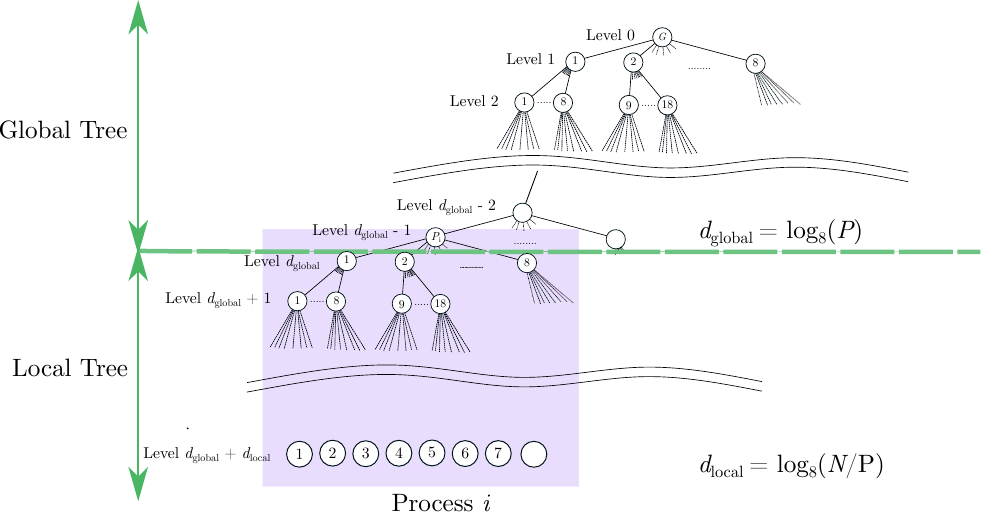}
  \caption{An illustration of octrees distributed over $P$ processes using Ibeid et al.'s specification \citep{ibeid2016performance}. Each node in the graph corresponds to a box, and each edge corresponds to a parent-child relationship, omitting some nodes for clarity. The root box of the global tree is labelled $G$, and corresponds to a box covering the problem domain. The root boxes of each local tree are labelled $P_i$ for $i \in [1, P]$ where $P$ is the total number of processes. Each node is labelled with its index when sorted in Morton order, at each level $l \in [0, d]$, where $d$ is the tree depth. Each process contains a non-overlapping subset of the leaf boxes}
  \label{fig:split_octree}
\end{figure}

Ibeid et al. \citep{ibeid2016performance} describe how the runtime communication complexity of the distributed \fmm can be reduced to

\begin{equation}
\label{eq:ibeid}
\bigO{\log{P} + \left(\frac{N}{P}\right)^{2/3}},
\end{equation}

which has been shown to apply to both uniform and adaptive cases \citep{yokota2014communication}. This reduction is achieved by splitting the \acrshort{hot} into two hierarchical components: (1) a \textit{global} or \textit{process} tree and (2) a \textit{local} tree within each process. The idea is illustrated in Figure~\ref{fig:split_octree} for the uniform case. In this scheme, each leaf of the global tree corresponds to a process, and serves as the root of a fine-grained local tree owned by that process. For adaptive trees, the leaves of the global tree are defined uniformly at the same level, while adaptivity is captured entirely within the local trees.

To understand how this splitting reduces redundant computation near the root, consider the upward pass. After each process computes the upward representation $\mathbsf{u}$ for the root of its local tree, the U2U (upward-to-upward) operation needs to be performed across local roots. However only one process needs to carry out this computation per group of eight processes – whose local roots correspond to sibling boxes (i.e. those sharing a common parent in the global tree). These processes can transmit their $\mathbsf{u}$ data to a single nominated process responsible for the U2U at that level. This logic can be applied recursively up the global tree, involving only $8^l$ active processes at level $l$, and yielding a total communication complexity of $\bigO{\log{P}}$, since each process participates in a constant number of communications and the height of the global tree can be seen to be $\log{P}$ from Figure~\ref{fig:split_octree}.

A similar pattern applies during the downward pass over the global tree. With reference to Figure~\ref{fig:split_octree}, for each level $l \in [2, d_{\text{global}} - 1]$, the $\mathbsf{u}$ data needed for the \textbf{VLI} step at a set of sibling boxes is determined by the interaction list $\mathcal{K}(\mathcal{C}(\mathcal{P}(\alpha)))$, where $\alpha$ is their shared parent. The nominated process, which already holds the $\mathbsf{u}$ data from the upward pass, can then broadcast or exchange with at most 26 processes whose data overlaps with this interaction list. Since each process communicates with a bounded number of neighbors, this pattern again yields a total communication cost of $\bigO{\log{P}}$ derived from the height of the global tree.

Communication for the downward pass within the local trees involves only $\bigO{1}$ neighboring processes. Regardless of whether the tree is uniform or adaptive, the $\mathbsf{u}$ and $\mathbsf{s}$ ghost data required for the local U, V, W, and X list interactions is restricted to boxes contained within, or overlapping with, those defined by $\mathcal{K}(\mathcal{C}(\mathcal{P}(\beta)))$, where $\beta$ is a box at levels $[d_{\text{global}}+1,\, d_{\text{global}} + d_{\text{local}}]$ within a local tree (see Figure~\ref{fig:split_octree}). These interactions remain confined to local trees rooted at boxes adjacent to the root of the current local tree, and therefore involve communication with at most 26 neighboring processes.

\section{Algorithm}\label{sec:algorithm}

Our algorithm iterates on the communication pattern described by \citep{ibeid2016performance}, with a specific focus on practical engineering concerns to ensure high-performance, portability by relying only on standard MPI collectives and ease of implementation in an existing single-node \fmm software.

We restrict ourselves to uniform trees, which simplifies the communication graph, and allows us to identify and resolve all data dependencies before any tree traversal. We delegate to a highly optimized \textbf{ULI} implementation to handle irregular point distributions.

\begin{enumerate}
  \item We create a static communication graph at setup time, and describe it using MPI's neighborhood communicators – introduced in MPI-3\footnote{Introduced in MPI-3, \textbf{neighborhood communicators} allow processes to define logical neighbors within a topology (e.g., a Cartesian grid or a graph-based partition tailored to an application), enabling efficient, localized communication with those neighbors. While the communicators themselves define logical relationships between \mpi processes, \mpi implementations can optimize communication by leveraging information about the physical hardware topology, when available, to improve performance. Furthermore, neighborhood communicators support localized variants of collective operations.}. As a result our implementation shares conceptual similarity with task-based runtime approaches.
  \item We evaluate the upward and downward passes on the global portion of the tree on \textit{a single process}, thus ensuring we maintain our crucial \textbf{VLI} optimizations developed for single-node performance, with the communication at the global level of the algorithm reduced to a single all-gather and all-scatter.
  \item Except for the non-uniform case at the very largest scales evaluating 32e9 particles, communication scaling is broadly supported by our complexity analysis.
\end{enumerate}

The description is split into two parts. Section \ref{sec:fmm:sub:algorithm_setup} describes the steps required to establish data dependencies between all boxes in the tree, and encode the static communication graph in Neighborhood communicators. Section \ref{sec:fmm:sub:algorithm} describes the runtime execution of our algorithm.

We note that neighbourhood communication is particularly well suited to the static, sparse communication graph implicit in Ibeid et al.'s scheme. Each process has only a limited number of neighbors with which it must exchange data: at most 26 during the local and global portions of the downward pass, and 8 during the global portion of the upward pass. Because these exchanges arise from adjacent local roots in the tree, the communication pattern is spatially localized in the application domain.

\subsection{Algorithm Setup – Precomputing Data Dependencies}\label{sec:fmm:sub:algorithm_setup}

\begin{figure}
  \includegraphics[height=0.86\textheight]{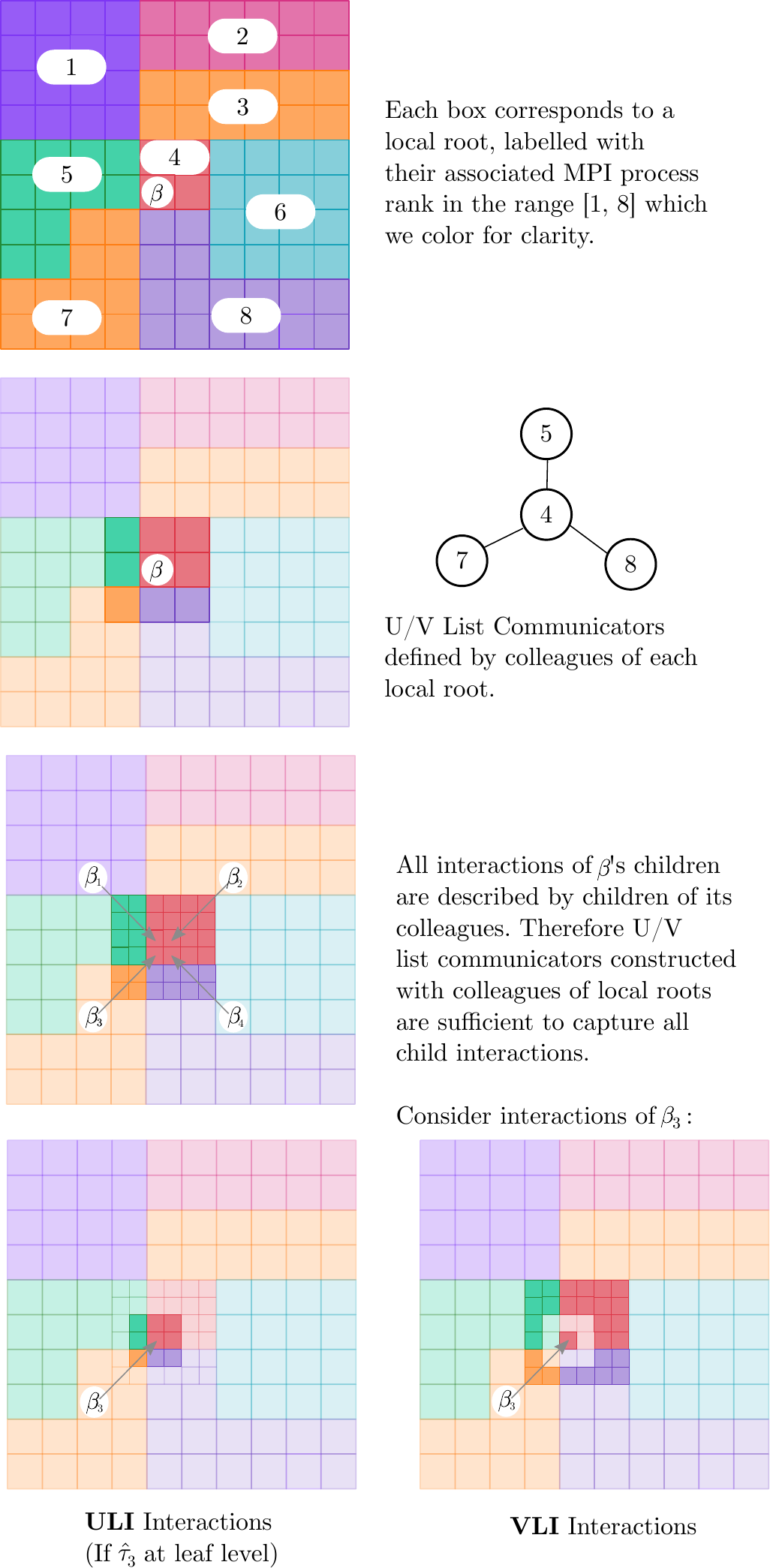}
    \caption{The first row illustrates an example of a layout of a distributed tree over 8 \mpi processes, each box corresponds to a local root, where we use a two dimensional tree for clarity. Colors correspond to \mpi processes, each of which can contain multiple local roots. In the second row we highlight a target box $\beta$ in process 4 and the edges and nodes that must be added to the neighborhood communication graphs at process 4 due to the interactions of $\beta$ and its descendants. These are calculated from the layout, interaction lists, and nearest neighbors of $\hat{\tau}$. In the third and fourth rows we consider the required communications for a child node of $\hat{\tau}$, showing how the neighborhood communication graph created for its parent can be re-used.}
  \label{fig:communicator_groups}
\end{figure}

\begin{mybox}[box:comm-setup]{Algorithm Setup}

\subsubsection*{\textbf{(1) Tree Construction}}

A user specifies the depth of the global tree $d_{\text{global}}$, which is used to encode all locally contained points within a process. These are then sorted with this coarse grained Morton key – the resulting leaves correspond to the \textitbf{local roots}. Each process, $i \in [1, P]$, can in principle be responsible for multiple local trees – a level of granularity that can be tuned by adjusting the splitters in the parallel sort.

Each process builds its own set of local trees, denoted \( T_{\text{local}}^{i,j} \), where \( j \in [1, \dots, n_P^i] \) indexes the local roots at process \( i \in [1,P]\). The \lettreefull of each local tree is constructed using \eqref{eq:let}, where boxes in the interaction list are defined from the halos that constitute the U and V interaction lists, those contained on ghost processes are checked for existence as we describe below.

\subsubsection*{\textbf{(2) Global Tree Layout}}

The set of local roots across all processes defines the structure of the global tree, which we refer to as the \textitbf{layout}. The layout maps \mpi ranks to the local roots they own. In the uniform case, the number of local roots assigned to a process acts as a proxy for computational load, enabling coarse-grained load balancing by adjusting the number of local trees assigned to each process.

To construct the layout, we perform an \texttt{MPI\_Allgatherv} over a global communicator to collect local root metadata (e.g., Morton keys and owning ranks) from all processes. Each process then has global knowledge of the layout.

\subsubsection*{\textbf{(3) Query Identification}}

For each box \( \beta \in T_{\text{local}}^{i,j} \), we have computed its U and V interaction lists in Step (1). To determine whether an interaction list box is locally available or must be fetched from a remote process, each process checks whether the box lies within the subdomain defined by the layout. Since all trees descend from known local roots, this check takes \(\bigO{P}\) per box. In the worst case, the total cost of forming these queries is \(\bigO{N}\) per process, but this can be reduced. In practice, a box's U and V interactions only involve trees rooted in the 26 neighboring subdomains. Thus, queries need only be formed for boxes potentially owned by these neighbors, reducing the cost to \(\bigO{26 \cdot (N/P)}\).

\subsubsection*{\textbf{(4) Forming Query Packets}}

Each process now constructs two query packets:

\begin{itemize}
    \item The \textbf{V-list query packet}: Morton keys and associated process ranks of all non-local boxes required for \textbf{VLI} interactions.
    \item The \textbf{U-list query packet}: Morton keys and associated process ranks of non-local boxes needed for \textbf{ULI} operations.
\end{itemize}

\subsubsection*{\textbf{(5) Building Neighborhood Communicators}}

We now construct neighborhood communicators, one for each of these queries which we call the \textbf{U-List Communicator} and \textbf{V-List Communicator}. We use \texttt{MPI\_Dist\_graph\_create} to encode inter-process dependencies identified in Step (4). Because interaction lists are spatially local in the problem domain, the resulting communication graphs are sparse. Each process communicates with at most 26 neighbors – the maximum number of adjacent subdomains in three dimensions. This worst case occurs when each process owns a single local root. If multiple local roots are assigned to the same process, the process-local communication graph is formed from the union of the corresponding local-root halos, so grouping roots onto one process does not increase the worst-case neighbor count beyond the single-root case. Figure~\ref{fig:communicator_groups} illustrates this graph construction for the communicators.

\subsubsection*{\textbf{(6) Establish Data Dependencies}}

\begin{enumerate}
    \item \textbf{Query Existence Exchange}: \\
    Each process calls \texttt{MPI\_Neighbor\_alltoallv} to send its U and V query packets to neighbors via their respective communicators. In the uniform case, each recipient checks whether the requested boxes exist locally (which can be accomplished in constant time if a set is used to store all boxes in the \lettree) and returns the keys of those that do, with another \texttt{MPI\_Neighbor\_alltoallv}.

    \item \textbf{U-list Data Exchange}: \\
    Using the U-list communicator, each process sends the source densities for ghost boxes needed for \textbf{ULI} interactions with a further call to \texttt{MPI\_Neighbor\_alltoallv}.

    \item \textbf{V-list Buffer Allocation}: \\
    Using the V-list existence responses, each process allocates buffers to receive the relevant $\mathbsf{u}$ vectors at runtime.
\end{enumerate}
\end{mybox}

Internally, each process stores its local box metadata using hash-based Morton key indexing, allowing \(\bigO{1}\) access to required data. We note that the after setup new source data can be added to existing trees by re-running the U-list data exchange, and clearing existing buffers.

\subsection{Algorithm – Runtime Execution}\label{sec:fmm:sub:algorithm}

Above we describe the process by which we establish data dependencies for the local trees, but not the global trees. A direct implementation of Ibeid et al.'s scheme \citep{ibeid2016performance} would require neighborhood communication among $8^l$ \mpi processes at level $l$ during the upward and downward passes over the global portion of the tree. At higher levels, the communicating processes become increasingly (logically) distant from one another, while each active process contributes only a single $\mathbsf{u}$ or $\mathbsf{d}$ vector.

Our approach both simplifies the communication steps, and allows us to compute $\mathbsf{u}$ and $\mathbsf{d}$ at all global tree boxes using our highly-optimized shared memory kernel implementations.

We do this by centralizing the evaluation of the global tree on a single process. We use \texttt{MPI\_Gatherv} to gather the $\mathbsf{u}$ vectors associated with all local roots on a nominated process, run the upward and downward passes on the global portion of the tree using the existing highly optimized shared-memory implementation, and return the $\mathbsf{d}$ vectors to their associated local roots using \texttt{MPI\_Scatterv}.

This introduces an artificial bottleneck, where the majority of the system is briefly idle while the global tree is traversed. However this cost is in practice small due to the efficiency of our shared memory implementation (especially without an expensive \textbf{ULI} step required). Additionally, the implementation of the communication for the global tree is reduced to just two global collectives.

Having performed the steps in Box \ref{box:comm-setup}, our runtime algorithm executes in \textitbf{local} and \textitbf{global} stages with reference to Figure \ref{fig:split_octree}:

\begin{mybox}[box:comm-runtime]{Runtime Execution}
\subsubsection*{\textbf{(1) Local Stage}}
\begin{enumerate}
    \item Run the upward pass for levels $l \in [d_{\text{global}}-1, d_{\text{global}}+ d_{\text{local}}]$ independently on all local trees, $T_{\text{local}}^{\,i, j}$ for $i \in [1, P]$ and $j \in [1, n_P^i]$ where $P$ is the total number of \mpi processes and $n_P^i$ is the number of local trees at process $i \in [1, P]$.
    \item Once the upward pass is complete, the $\mathbsf{u}$ data required during the downward pass is available for all local trees via the V-list communicator. This data is exchanged using \texttt{MPI\_Neighbor\_alltoallv} with the V-list communicator and inserted into the buffers, both created during setup.
\end{enumerate}
\subsubsection*{\textbf{(2) Global Stage}}
\begin{enumerate}
    \item Select a nominated \mpi process to manage the global tree computation. Gather $\mathbsf{u}$ data associated with each local root to the nominated process using \texttt{MPI\_Gatherv} over the whole network.
    \item On the nominated process, construct an explicit global tree, $T_{\text{global}}$, and perform the upward and downward passes entirely in \textit{shared memory} (excluding the S2T operation) for levels $l \in [0, d_{\text{global}}-1]$.
    \item Scatter the $\mathbsf{d}$ vectors at the leaf level of the global tree, $T_{\text{global}}$, back to their associated processes using the layout, calculated during algorithm setup, with \texttt{MPI\_Scatterv} over the whole network. These expansions correspond to the local roots of each local tree, $T_{\text{local}}^{\,i, j}$.
    \item Each \mpi process can now perform the downward pass for its local trees, $T_{\text{local}}^{\,i, j}$, independently in parallel for levels $l \in [ d_{\text{global}}-1, d_{\text{global}}+ d_{\text{local}}]$, with no further communication required.
\end{enumerate}
\end{mybox}

\section{Complexity Analysis}\label{sec:analysis}

\begin{figure}
  \centering
  \includegraphics[width=0.7\textwidth]{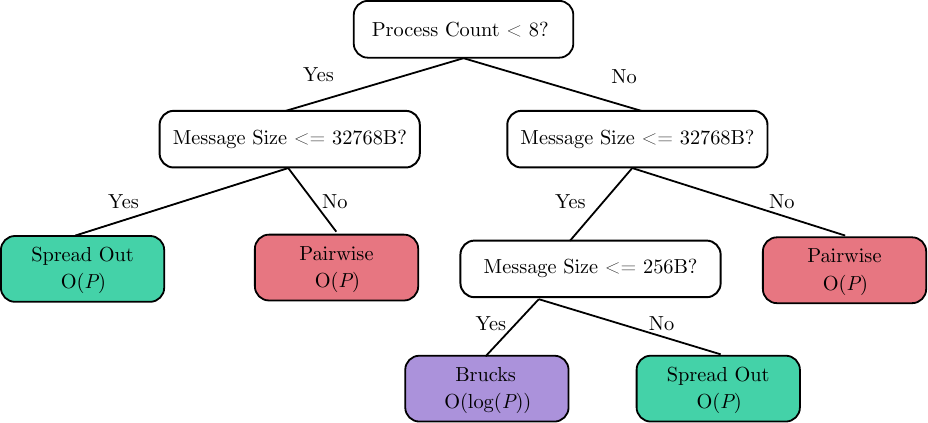}
  \caption{The default decision tree in MPICH for choosing an algorithm for \texttt{MPI\_Alltoallv}. Adapted from Figure 2 in \citep{Netterville2022}.}
  \label{fig:mpich_decision_tree}
\end{figure}

Our goal in this section is not to argue that asymptotic complexity alone determines the merit of the method, but rather to show that the communication structure proposed above remains compatible with favorable runtime communication bounds in the regimes targeted by our implementation. We therefore also account for the behavior of modern \mpi implementations, which commonly use heuristics based on the number of \mpi processes and the size of messages to select an appropriate communication algorithm. For example, Figure~\ref{fig:mpich_decision_tree} illustrates the decision tree used by MPICH for \texttt{MPI\_Alltoallv}; other implementations use similar strategies.

\subsubsection{Algorithm Setup}

The tree construction during Step (1) in Box \ref{box:comm-setup} is dictated by the communication complexity of the parallel sort, which we described in Section \ref{sec:fmm:sub:distributed}. We treat this sorting phase as a preprocessing step, separate from the distributed \fmm phase analyzed below, and in this work we use Samplesort for simplicity of implementation rather than an asymptotically more specialized alternative.

Step (2) in Box \ref{box:comm-setup} constructs the layout - associating all local roots with \mpi ranks and storing at each process. This involves a call to \texttt{MPI\_Allgatherv} over the entire network. Each process contributes a small message: the metadata of its local roots, typically Morton keys and ranks, which can be encoded using 64-bit integers. If the total message size is less than $\sim 10$~KB per-process, which corresponds to over 1000 local roots per process, modern \mpi implementations can use logarithmic algorithms such as Bruck's algorithm by default, resulting in a communication complexity of \( \bigO{\log P} \) for this step \citep{Thakur2005}.

During step (6) in Box \ref{box:comm-setup}, sub-steps (1) and (2) involve neighborhood exchanges of query packets and ghost data between each process and its spatial neighbors. In the worst case this is bounded by 26 neighboring processes in three dimensions, and does not grow with \( P \). The resulting communication complexity therefore matches that of the local portion of the downward pass described in Sections 5.1.3 and 5.1.4 of \citep{ibeid2016performance}, given by

\[
    \bigO{(N/P)^{2/3}}.
\]

\subsubsection{Algorithm}

\subsubsection{Local Portion}

Step (2) during the local portion of Box \ref{box:comm-runtime} again requires neighborhood communications between $\bigO{1}$ \mpi processes for each process as during sub-steps (1) and (2) during step (6) of Box \ref{box:comm-setup}, resulting in the same complexity estimate.

\subsubsection{Global Portion}

Communication during the global portion in Box \ref{box:comm-runtime} occurs in steps (1) and (3): a gather of \( \mathbsf{u} \) vectors and a scatter of \( \mathbsf{d} \) vectors associated with local roots. These steps are performed using \texttt{MPI\_Gatherv} and \texttt{MPI\_Scatterv}, respectively. As mentioned above, when message sizes are small \mpi implementations default to logarithmic complexity communication schemes \( \bigO{\log P} \). However, if message sizes exceed implementation thresholds, fallback algorithms with \( \bigO{P} \) complexity may be triggered. The message size per process depends on the number of local roots \( n_P^i \), the floating-point precision \( b \in \{32, 64\} \), and the \fmm `expansion order' \( \tilde{P} \) – which tunes the accuracy of the evaluated field. This is of $\bigO{\tilde{P}^2}$ in 3D FMMs, and in the \kifmm implementation used by our software \citep{ying2004kernel}, each \( \mathbsf{u} \) or \( \mathbsf{d} \) vector has length \( 6(\tilde{P}-1)^2 + 2 \).

The total message size per process is thus approximately

\[
    \text{Message size} \approx n_P^i \cdot (6(\tilde{P}-1)^2 + 2) \cdot \frac{b}{8} \text{ bytes}.
\]

\subsubsection{Runtime Communication Bound}

Assuming message sizes remain below threshold and spatial communication graphs remain sparse, the runtime communication complexity is
\begin{equation}
    \label{eq:chpt:field_translation:sec:distributed:sub:analysis:complexity}
    \bigO{\log P + (N/P)^{\,2/3}},
\end{equation}
which matches the best-case bound of Ibeid et al. \citep{ibeid2016performance}. We emphasize, however, that this result should be interpreted as support for the practicality of the design rather than its sole motivation: the main benefit of the method remains that it realizes hierarchical communication using only standard collectives and a precomputed communication layout while preserving the high-performance shared-memory kernels used locally.

Achieving this bound in practice depends on the underlying hardware and software stack. In particular, interconnect technologies must support hierarchical collectives, and message sizes must be tuned to remain below system-specific thresholds. If not, global communication may exhibit linear complexity, resulting in an upper bound of

\[
    \bigO{P + (N/P)^{\,2/3}}.
\]

\section{Benchmarks}\label{sec:benchmarks}

The ARCHER2 supercomputer is a HPE Cray EX system, consisting of 5680 compute nodes each containing two AMD EPYC 7742 64-core 2.25GHz \cpus, networked with a HPE Cray Slingshot interconnect in a Dragonfly topology. There are 8 NUMA regions per node, with 16 cores per NUMA region and up to 512 GB of memory per node, or equivalently 4 GB per core. In total there are 750080 \cpu cores, supporting 1500160 software threads with hyper-threading. Each node consists of two \cpu sockets, each of which contains 8 \acrfullpl{ccd}; each \acrshort{ccd} consists of 2 \acrfullpl{ccx}, and each \acrshort{ccx} owns 4 compute cores that share 16 MB of L3 cache. This hierarchy makes process placement an important part of performance tuning for our implementation, since the local portion of the \fmm should remain within a shared cache or NUMA region whenever possible. We note that users are able to request at most 512 nodes of the system.

We benchmark our software on two point distributions: a uniform random distribution in the volume, and points distributed randomly on the surface of a unit sphere. The spherical distribution is intentionally more challenging for a uniform tree, since many fine boxes are empty and the work associated with the local trees is therefore less balanced than in the uniform case. All results reported here use expansion order \(\tilde{P}=3\) in single precision for the 3D Laplace problem \eqref{eq:laplace_kernel}. Each benchmark configuration is repeated five times, and the runtime and efficiency plots report mean wall times together with bands showing one standard deviation across runs. Some run-to-run variability on ARCHER2 is expected even for a fixed configuration, since batch-scheduled runs exhibit system-level variability that is not directly controlled by the application. In the non-uniform case this system-level variability is compounded by the intrinsic imbalance of the workload and messages sent across the network, so the derived parallel-efficiency curves are more variable and we use the runtime curves as the primary indicator of scaling behavior. These experiments are designed to test three claims of the method: first, that preserving shared-memory locality through an appropriate rank granularity matters strongly for performance; second, that a communication scheme based on standard collectives and a precomputed layout still delivers good large-scale runtime performance for the distributed \fmm phase; and third, that the centralized global phase remains a secondary cost in practice.

To study the effect of node-level granularity we compare three \mpi process placement strategies on ARCHER2:
\begin{enumerate}
  \item \textbf{CCX pinning}: each \mpi process is pinned to a single \acrshort{ccx}, corresponding to 4 \cpu cores that share one L3 cache.
  \item \textbf{CCD pinning}: each \mpi process is pinned to a \acrshort{ccd}, corresponding to 8 \cpu cores, or two \acrshortpl{ccx}, within a single NUMA region.
  \item \textbf{Socket pinning}: each \mpi process is pinned to an entire socket, corresponding to one AMD EPYC 7742 64-core \cpu.
\end{enumerate}

At 4e9 points we compare all three strategies directly. Although the algorithm can in principle assign multiple local trees to a rank, the reported ARCHER2 weak-scaling runs use one local tree per \mpi rank. The benchmark configurations are chosen so that each node carries the same total number of points despite the differing rank granularity: CCX pinning uses 32 ranks per node with 250000 points per rank, CCD pinning uses 16 ranks per node with 500000 points per rank, and socket pinning uses 2 ranks per node with $4\times10^6$ points per rank. In all three cases this gives $8\times10^6$ points per node, so the comparison is made at matched node-level problem size even though the amount of work owned by each rank differs substantially across strategies. These results should therefore be interpreted as a comparison of complete node-level configurations rather than of hardware pinning in isolation. Within each pinning strategy, the total number of ranks then increases in factors of eight, so the global tree depth increments by one at each scaling step. The local depth is fixed within a given benchmark configuration, but chosen per strategy to keep the local trees at a reasonable resolution in terms of estimated points per leaf box: the 4e9 campaigns use local depth 4 for CCX and CCD pinning, and local depth 5 for socket pinning. Since all benchmark results use \(\tilde{P}=3\) in single precision, each rank contributes a single local-root \( \mathbsf{u} \) or \( \mathbsf{d} \) vector of 26 floating-point values, corresponding to 104 bytes, so the global gather/scatter messages remain firmly in the small-message regime assumed in Section~\ref{sec:analysis}. For the largest 3.2e10-point experiments we report results for socket pinning, giving one \mpi rank per socket and 1024 ranks on 512 nodes. These socket-pinned campaigns use local depth 5 for the uniform case and 6 for the spherical case, again reflecting a choice of local tree resolution rather than a change in the weak-scaling construction itself. This choice is deliberate: for the ARCHER2 weak-scaling results reported below, socket pinning provides the most robust large-scale behavior among the node-level configurations we study because it leaves a larger shared-memory workload within each rank and fewer ranks to coordinate during global collective phases, even though finer-grained placements remain useful when strong scaling fixed problem sizes.

\subsection{Weak Scaling at 4e9 Points}

\begin{figure}
  \centering
  \includegraphics[width=\textwidth, trim=0 0 0 1.1cm, clip]{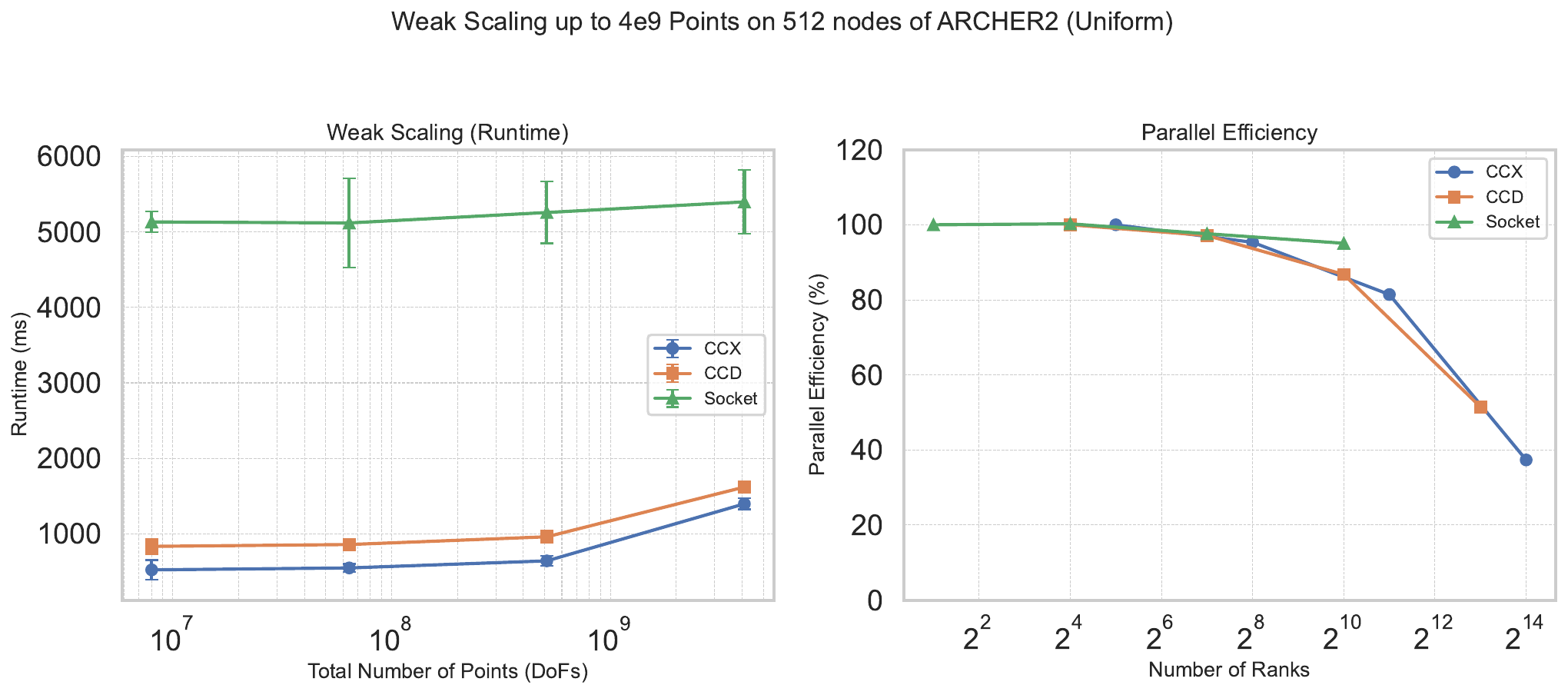}\\[0.8em]
  \includegraphics[width=\textwidth, trim=0 0 0 1.1cm, clip]{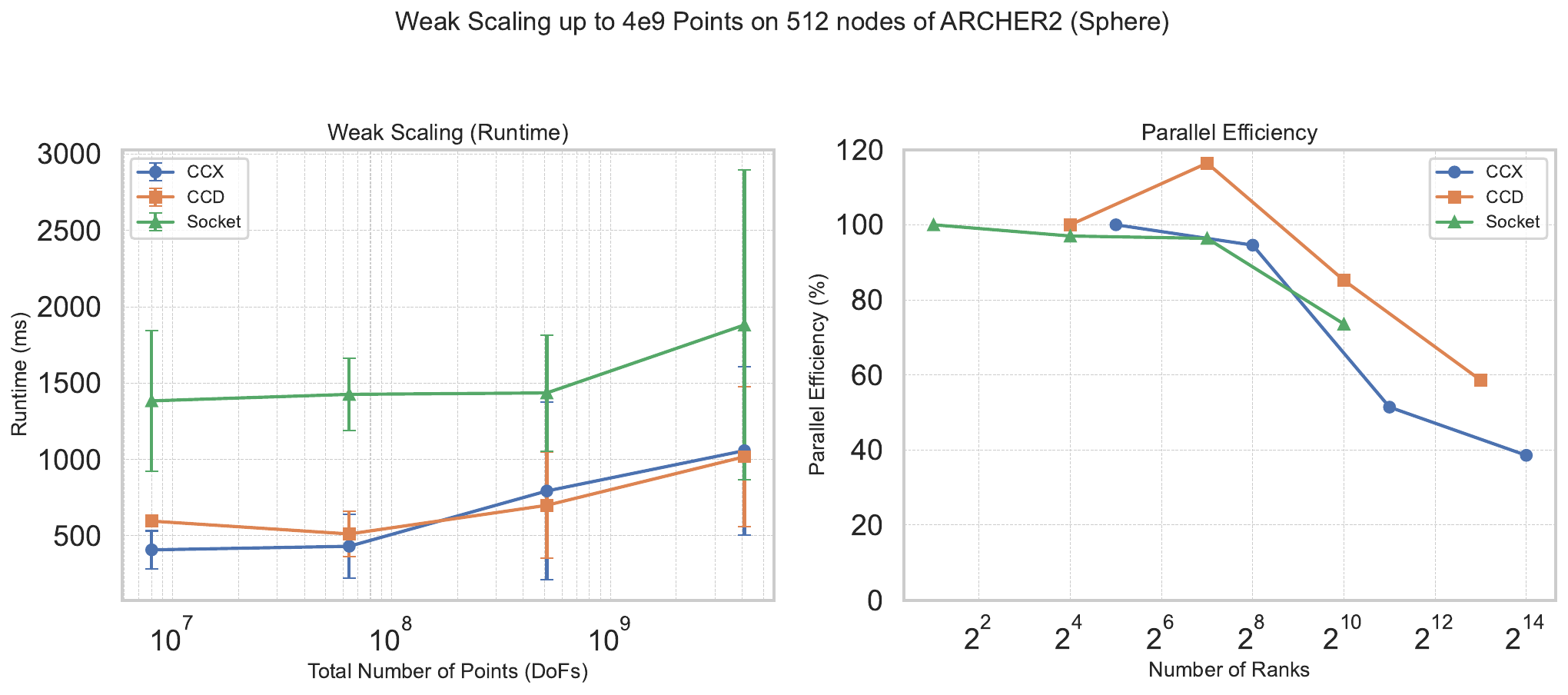}
  \caption{Weak scaling to 4e9 points on 512 ARCHER2 nodes. Top: uniform random distribution. Bottom: spherical surface distribution. Blue circles, orange squares, and green triangles correspond to CCX, CCD, and socket pinning, respectively. Times are reported from wall time, error bars are reported from standard deviation over 5 runs.}
  \label{fig:weak_scaling_4e9}
\end{figure}

Figure~\ref{fig:weak_scaling_4e9} compares weak scaling up to 4e9 points on 512 ARCHER2 nodes for all three pinning strategies using the hierarchy-aware construction described above. In these experiments, runtime is reported from wall time. These experiments expose the trade-off between fine-grained \mpi parallelism and reuse of shared-memory locality inside a rank. Across both distributions we observe useful weak scaling over the full machine range, with distributed \fmm runtime remaining in the low single-digit seconds even at the largest runs shown here. The spherical distribution exhibits visibly larger error bars and more irregular efficiency curves, reflecting a combination of workload imbalance from the non-uniform point set and the run-to-run system noise already noted above, so we interpret the efficiency curves in conjunction with the runtime data rather than in isolation. The dominant costs between each process pinning strategy appear to be determined by the number of points-per-MPI process, determining the depth of the local trees to traverse within each rank.

The socket pinning regime appears to have the best parallel efficiency. However, this is likely a reflection of the performance of our global collectives which now coordinate fewer ranks for the same problem size. However, this also indicates that this is a regime we can use to scale to significantly larger problem sizes – if we accept the larger fixed costs per-rank with each rank being responsible for a deeper \fmm tree (containing more points). This cost is acceptable if the shared-memory implementation of the \fmm is sufficiently optimized – reflected in our low overall runtimes.

\subsection{Weak Scaling at 3.2e10 Points}

\begin{figure}
  \centering
  \includegraphics[width=\textwidth]{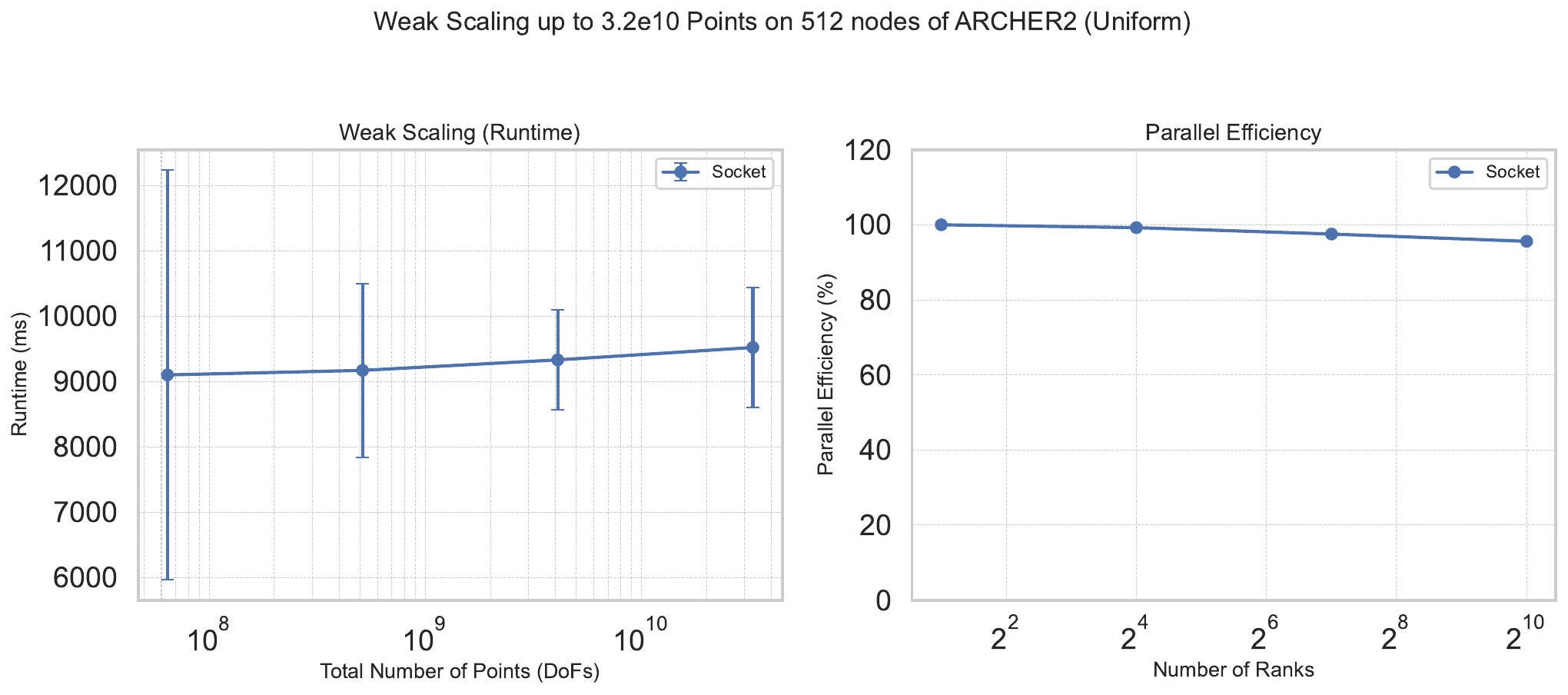}\\[0.8em]
  \includegraphics[width=\textwidth]{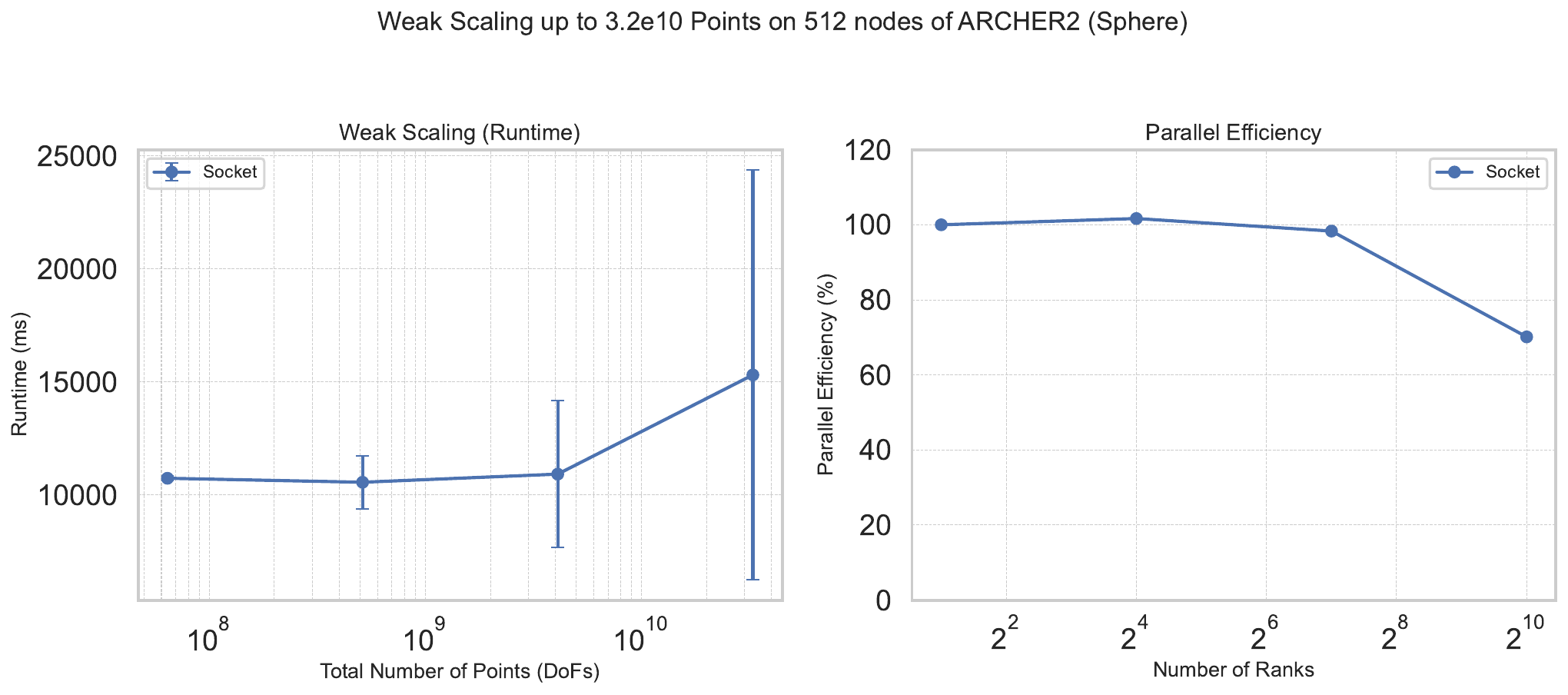}
  \caption{Weak scaling to 3.2e10 points on 512 ARCHER2 nodes using socket pinning. Top: uniform random distribution. Bottom: spherical surface distribution.}
  \label{fig:weak_scaling_3p2e10}
\end{figure}

Figure~\ref{fig:weak_scaling_3p2e10} shows the socket-pinned weak-scaling construction extended to the largest experiments. Each rank owns a single local tree, responsible for 32e6 points each, each local tree is refined to depth 5 in the uniform case, and depth 6 in the spherical case. We require only $2^{10}$ ranks due to the large numbers of points handled per-rank.

This figure is the clearest demonstration of the regime we ultimately target: a coarse-grained distributed decomposition in which each rank still owns enough local work for the shared-memory kernels to remain the main determinant of runtime.

The large error bars for given rank configurations are hard to attribute to a single cause from our data. We have little control over the exact rank placement, or noise on the system due to co-running programs. However, just looking at the mean runtimes we observe good parallel efficiency, except for the most extreme case for spherical distributions where the load imbalance due to our imposition of uniform trees becomes visible despite the efficiency of our shared memory kernels.

There clearly exists a trade-off between regimes where our communication strategy begins to degrade in efficiency and the maximum number of particles per-rank our shared memory implementation can handle. However, even in the least favorable cases with up to 3.2e10 non-uniform points exposing a significant load-imbalance in work per-rank, as well as the messages sent across the network, our observed mean wall times remain in the region of 10s of seconds.

\subsection{Strong Scaling}
\begin{figure}
  \centering
  \includegraphics[width=\textwidth, trim=0 0 0 1.1cm, clip]{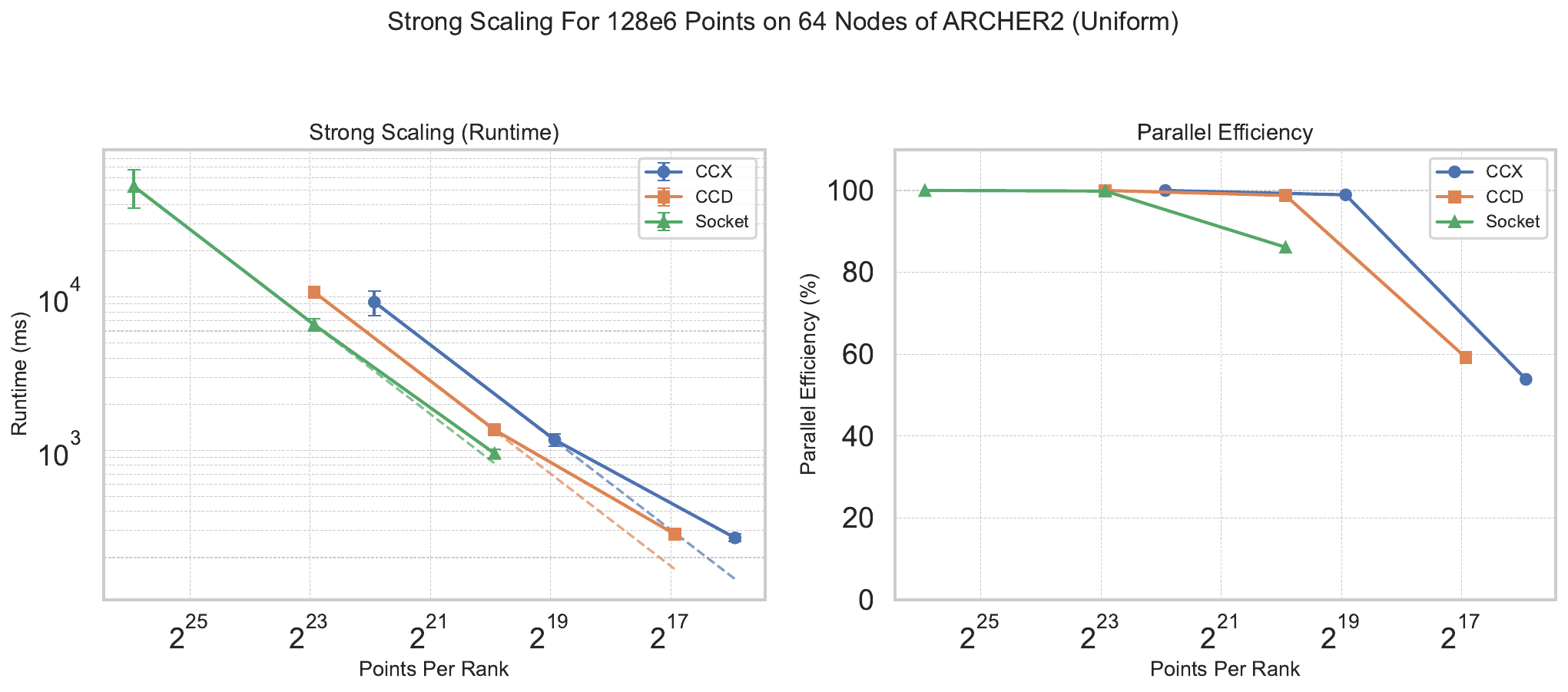}\\[0.8em]
  \includegraphics[width=\textwidth, trim=0 0 0 1.1cm, clip]{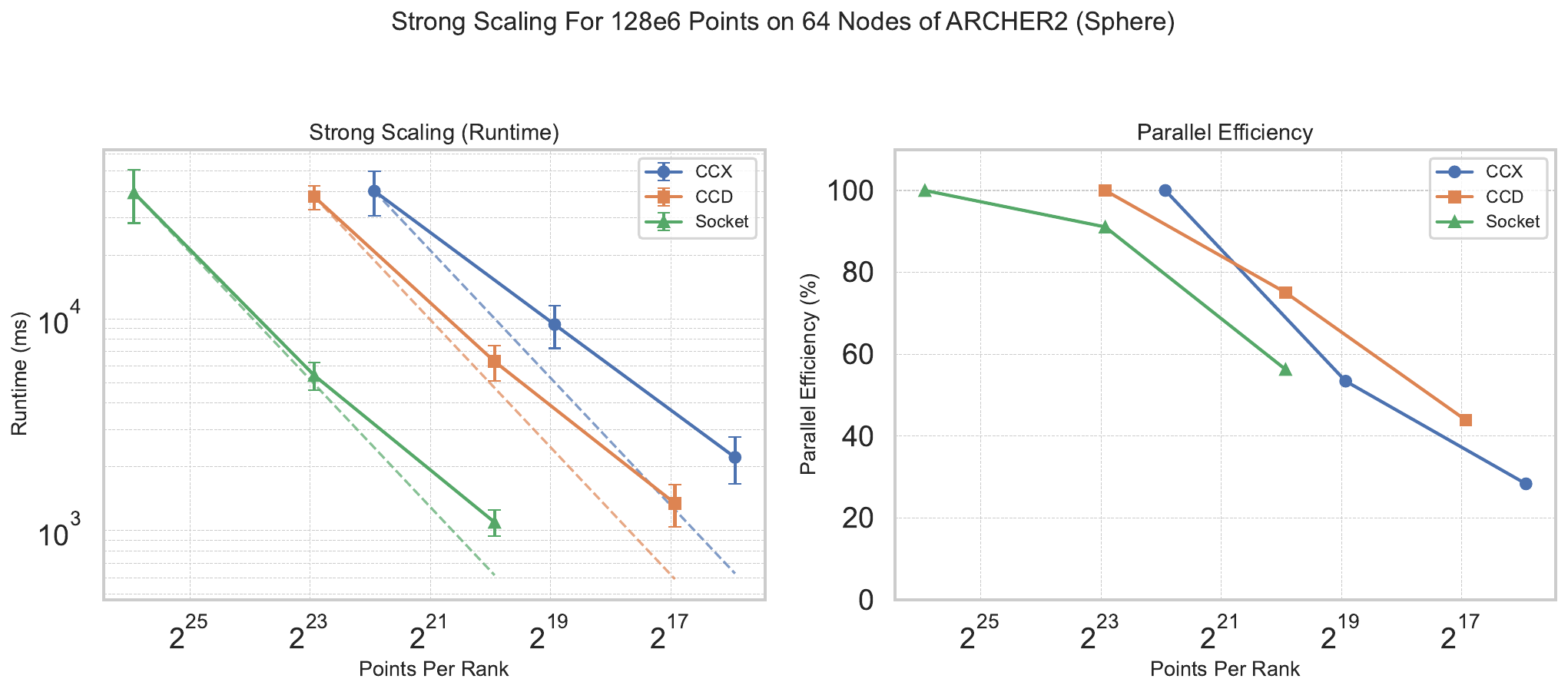}
  \caption{Practical strong-scaling study for a fixed problem size of $128 \times 10^6$ points, using up to 64 ARCHER2 nodes. Top: uniform random distribution. Bottom: spherical surface distribution. Blue circles, orange squares, and green triangles correspond to CCX, CCD, and socket pinning, respectively. Dashed lines indicate ideal strong-scaling trends for the corresponding pinning strategy and should be read only as visual references. The local depth is adjusted across the sequence to maintain comparable estimated leaf occupancy, so the tree is not identical at every point.}
  \label{fig:strong_scaling}
\end{figure}

Figure~\ref{fig:strong_scaling} reports strong-scaling experiments for a fixed problem of $128 \times 10^6$ points, using up to 64 ARCHER2 nodes. In contrast to the weak-scaling study, the total number of points is held fixed here while the resources are increased in factors of two. Within each pinning strategy the number of ranks per node is fixed by the placement choice, so doubling the node count also doubles the total number of ranks and halves the points owned by each rank. The local depth is then adjusted so that the estimated leaf occupancy remains comparable across the sequence. The resulting figure should therefore be interpreted as a practical strong-scaling study for this implementation, rather than as a strictly fixed-discretization scaling test with an identical tree at every point. In particular, the reference lines should be read only as visual guides, since the tree structure is mildly adjusted across the sequence. These runs nevertheless complement the weak-scaling results by showing how far each process placement strategy can be pushed before the amount of work per rank becomes too small to maintain good efficiency. In this setting the picture is different from the largest weak-scaling runs: CCX and CCD pinning remain efficient to finer granularities than socket pinning, particularly for the uniform distribution, because they expose efficient process-level parallelism once the per-rank work has become small – ;likely due to the efficiency of our shared-memory kernels optimized for these configurations.

For the spherical distribution, efficiency drops earlier for all three strategies, consistent with the greater imbalance already seen in the weak-scaling runs. Together, the weak- and strong-scaling results therefore support a consistent interpretation: for the largest production-style weak-scaling runs on ARCHER2, socket pinning is the most effective granularity because it preserves a larger shared-memory workload per rank, while finer-grained pinning is most useful when aggressively strong scaling a fixed problem.

\subsection{Setup Breakdown}

\begin{figure}
  \centering
  \includegraphics[width=\textwidth]{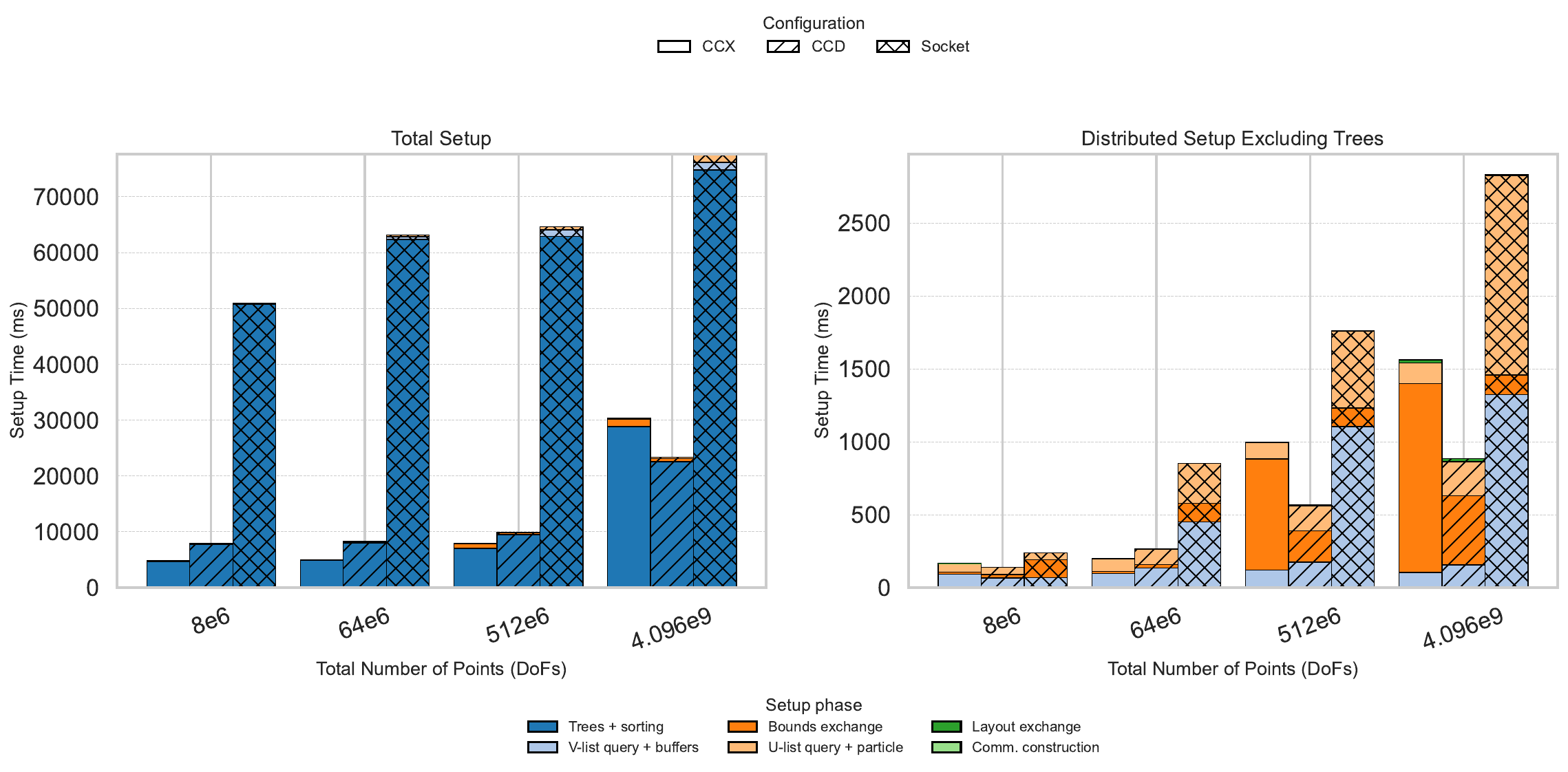}\\[0.8em]
  \includegraphics[width=\textwidth]{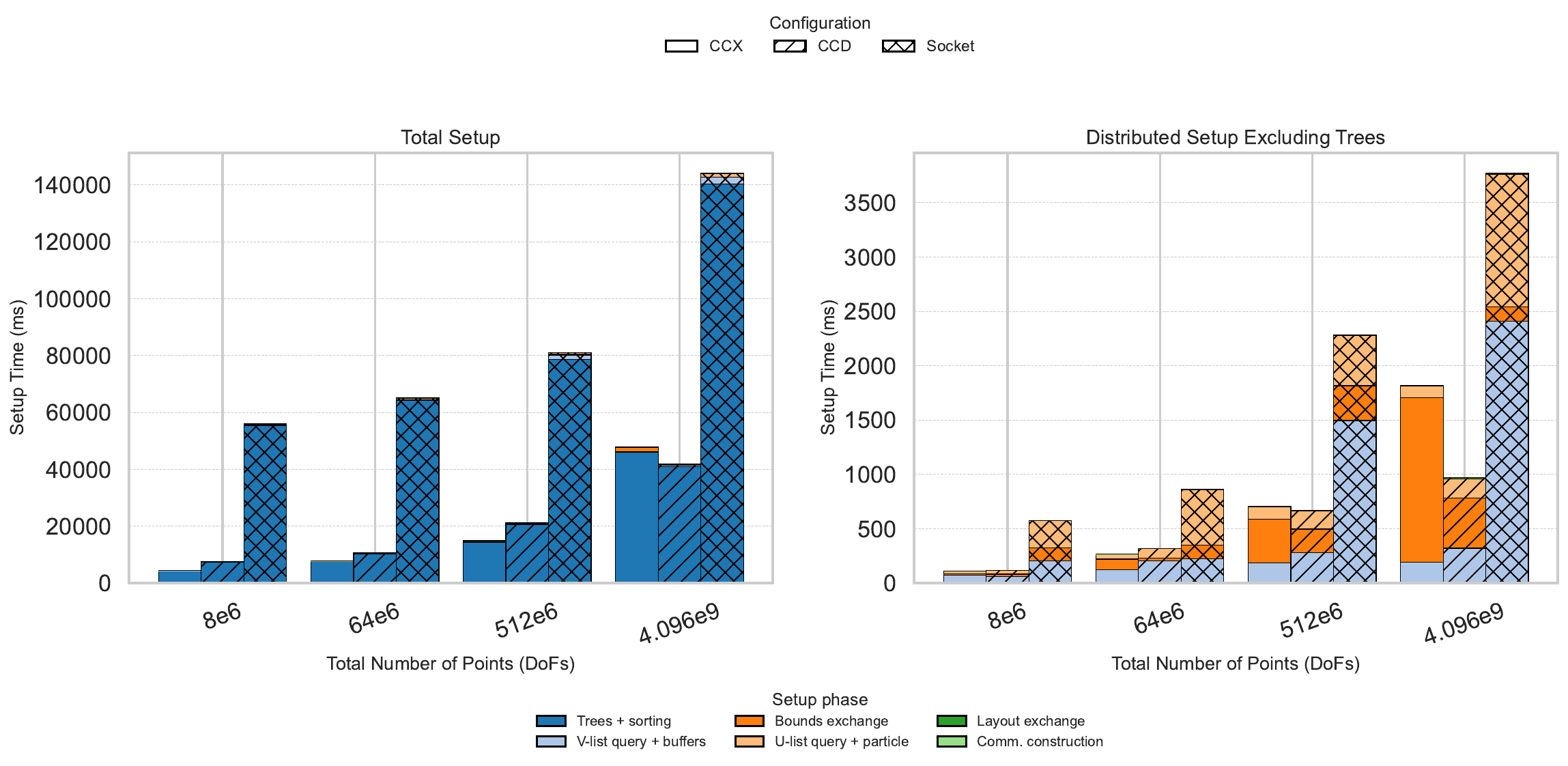}
  \caption{Setup-time breakdown for the socket-pinned weak-scaling configurations on ARCHER2 at 4e9 points. Top: uniform random distribution. Bottom: spherical surface distribution. In each plot, the left panel shows the total setup cost including tree construction and the initial distributed sort, while the right panel excludes that dominant phase to reveal the remaining distributed setup stages: domain-bounds exchange, global layout exchange, communicator construction, U-list query and particle exchange, and V-list query and buffer setup.}
  \label{fig:setup_breakdown}
\end{figure}

\begin{figure}
  \centering
  \includegraphics[width=\textwidth]{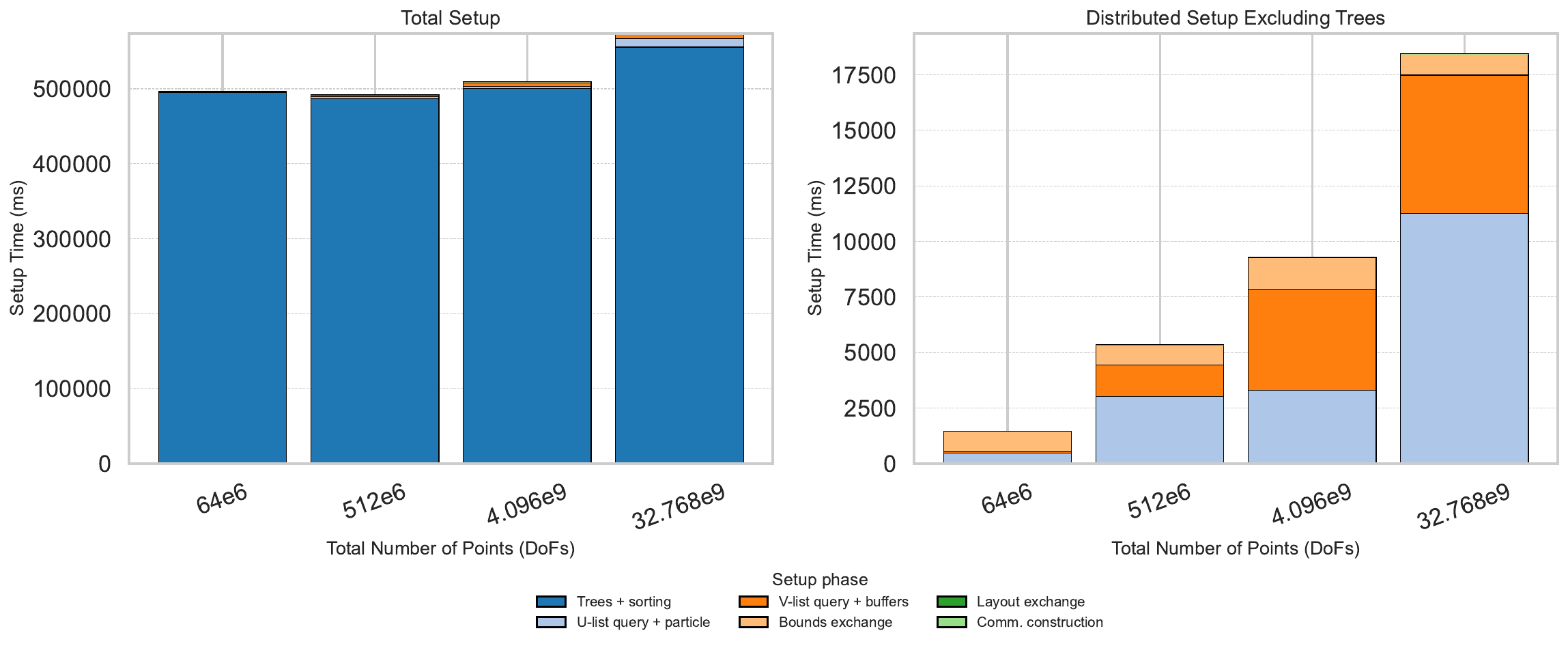}\\[0.8em]
  \includegraphics[width=\textwidth]{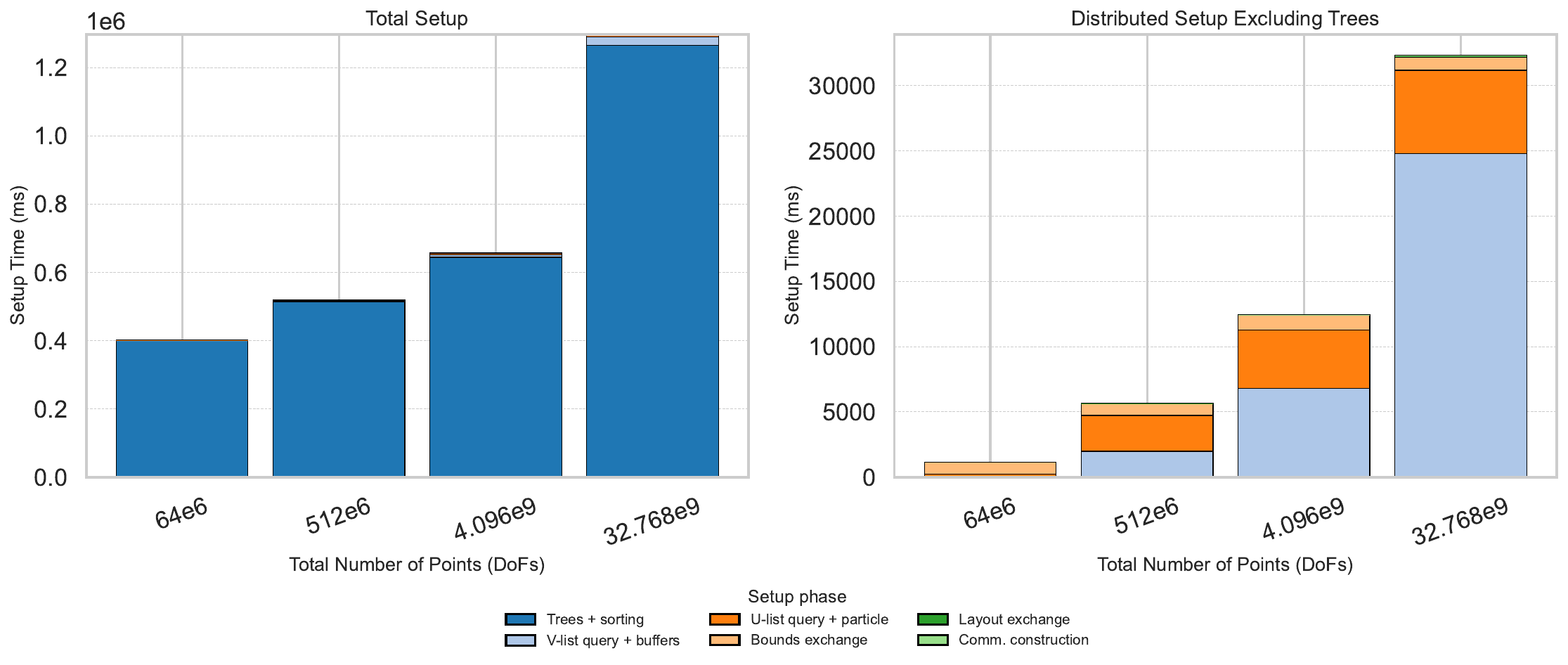}
  \caption{Setup-time breakdown for the largest socket-pinned weak-scaling configurations on ARCHER2, up to $3.2\times 10^{10}$ points. Top: uniform random distribution. Bottom: spherical surface distribution. As in Figure~\ref{fig:setup_breakdown}, the left panel shows the total setup cost including tree construction and the initial distributed sort, while the right panel excludes that dominant phase to expose the remaining distributed setup stages.}
  \label{fig:setup_breakdown_big}
\end{figure}

Figure~\ref{fig:setup_breakdown} decomposes the one-off setup cost for the socket-pinned weak-scaling configurations at 4e9 points, while Figure~\ref{fig:setup_breakdown_big} shows the same breakdown for the largest runs up to 3.2e10 points. In both settings, the setup is dominated by tree construction together with the initial distributed sort, so each combined plot also reports a second panel in which this dominant phase is excluded to make the remaining setup stages visible. These right-hand panels isolate the communication-specific setup work discussed in Section~\ref{sec:fmm:sub:algorithm_setup}: the domain-bounds exchanges used to establish box bounds, the global layout exchange, communicator construction, and the U-list and V-list query/setup phases.

\subsection{Runtime Communication Breakdown}

\begin{figure}
  \centering
  \includegraphics[width=\textwidth]{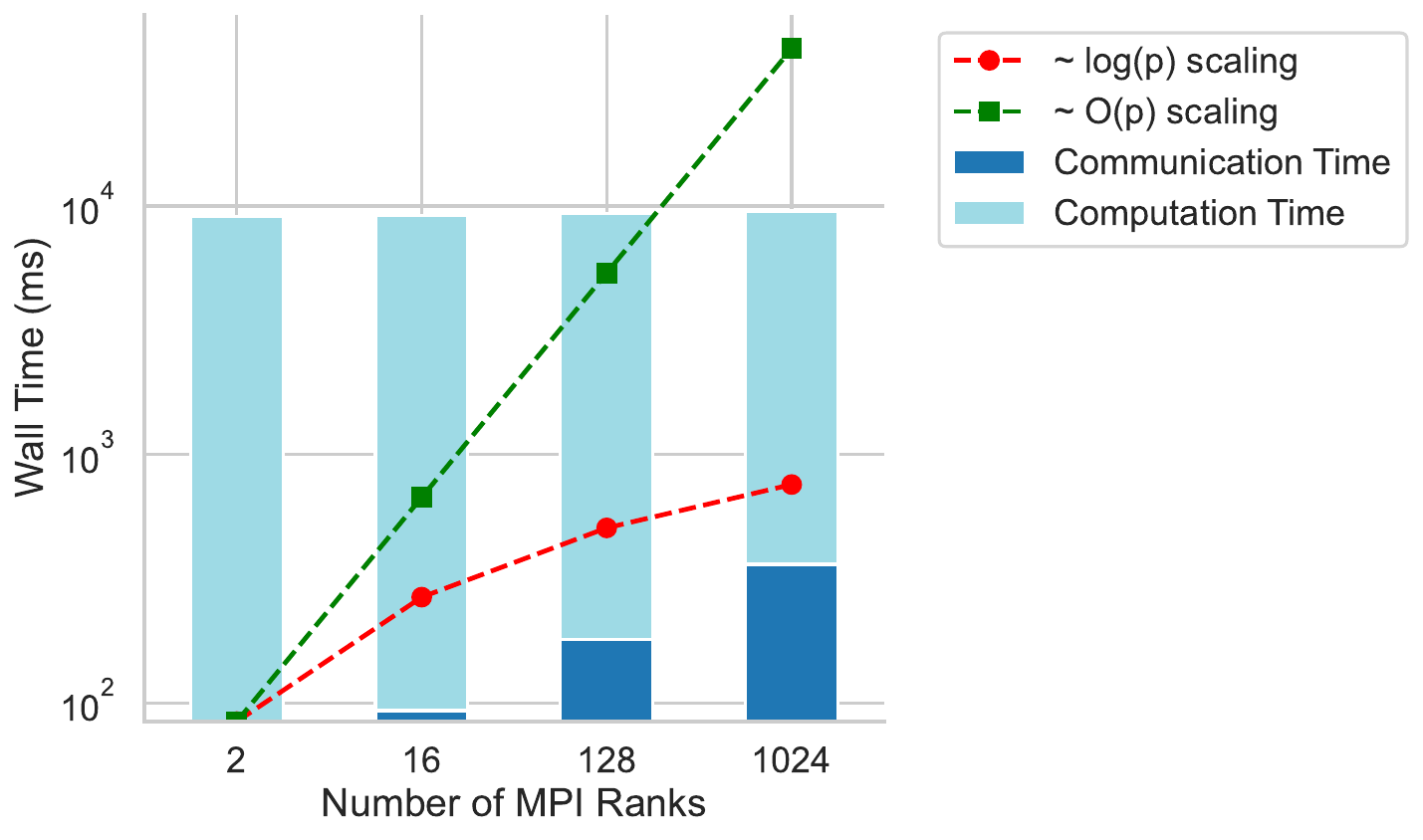}\\[0.8em]
  \includegraphics[width=\textwidth]{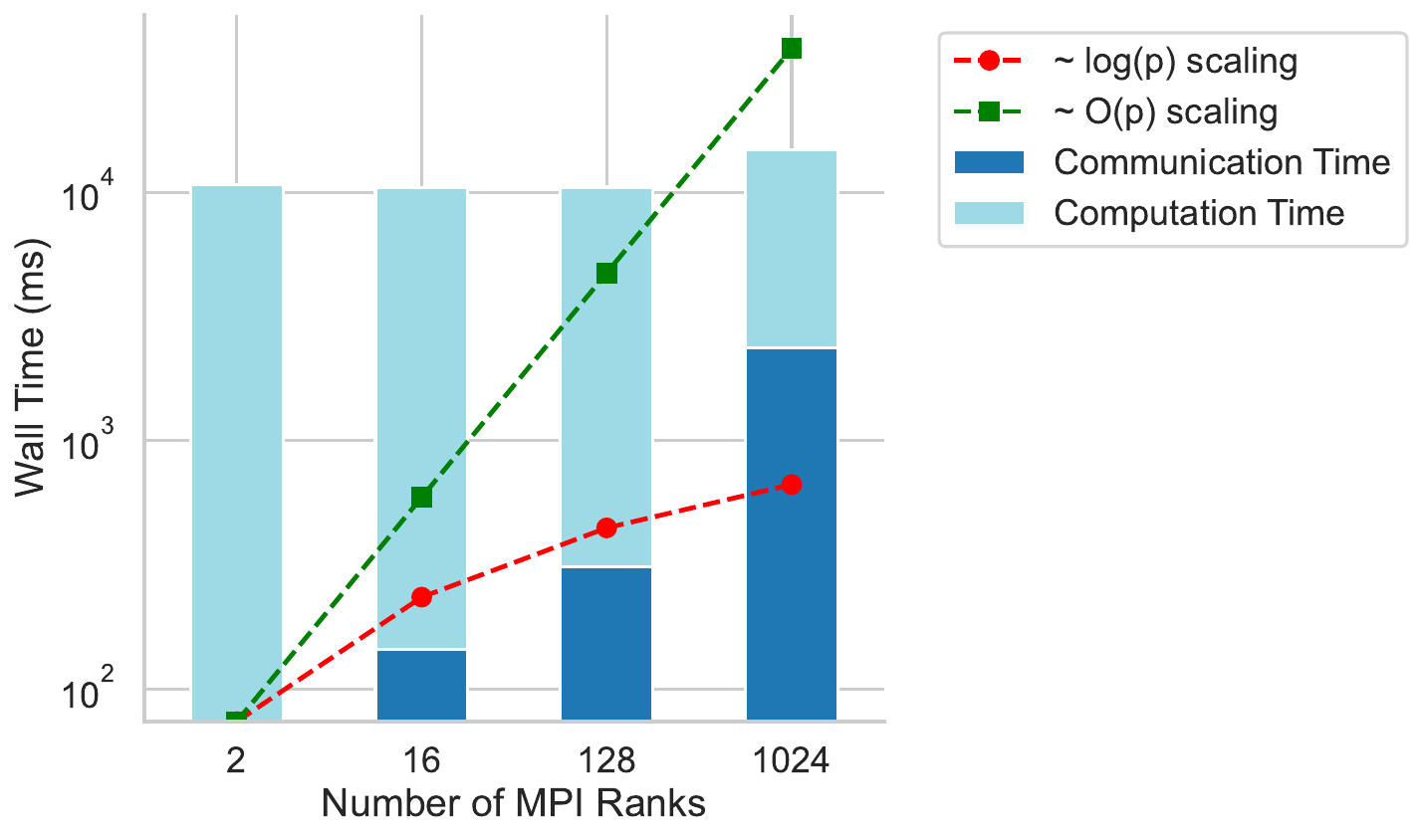}
  \caption{Aggregate communication and computation time for the same socket-pinned weak-scaling configurations used in the preceding scaling study. Top: uniform random distribution. Bottom: spherical surface distribution. The stacked bars report mean times over five repeated runs and separate communication from computation, while the dashed red and green curves indicate reference \(\log P\) and \(P\) trends computed from the ranks in each figure and normalized with respect to the first measured value.}
  \label{fig:comm_trend}
\end{figure}

\begin{figure}
  \centering
  \includegraphics[width=\textwidth]{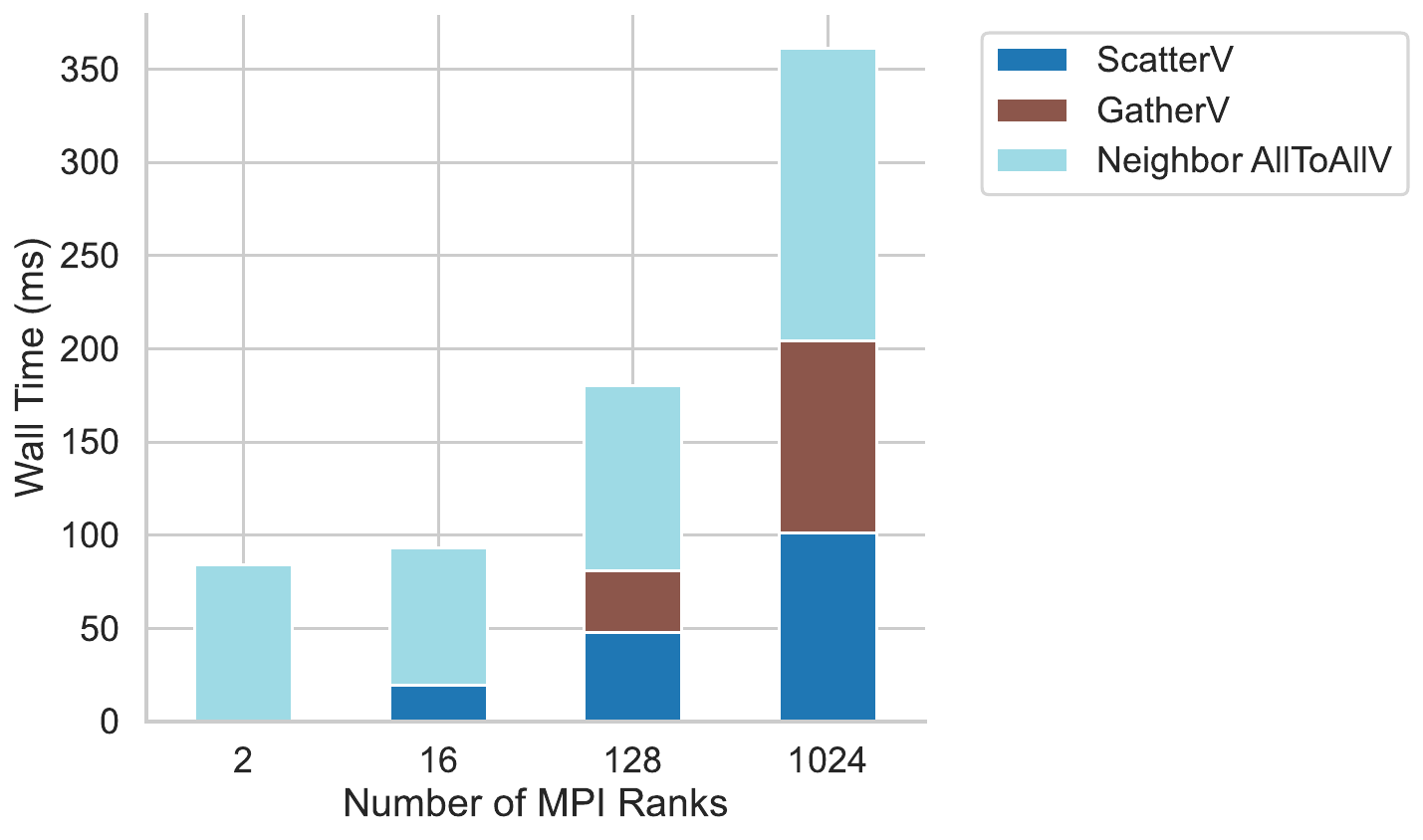}\\[0.8em]
  \includegraphics[width=\textwidth]{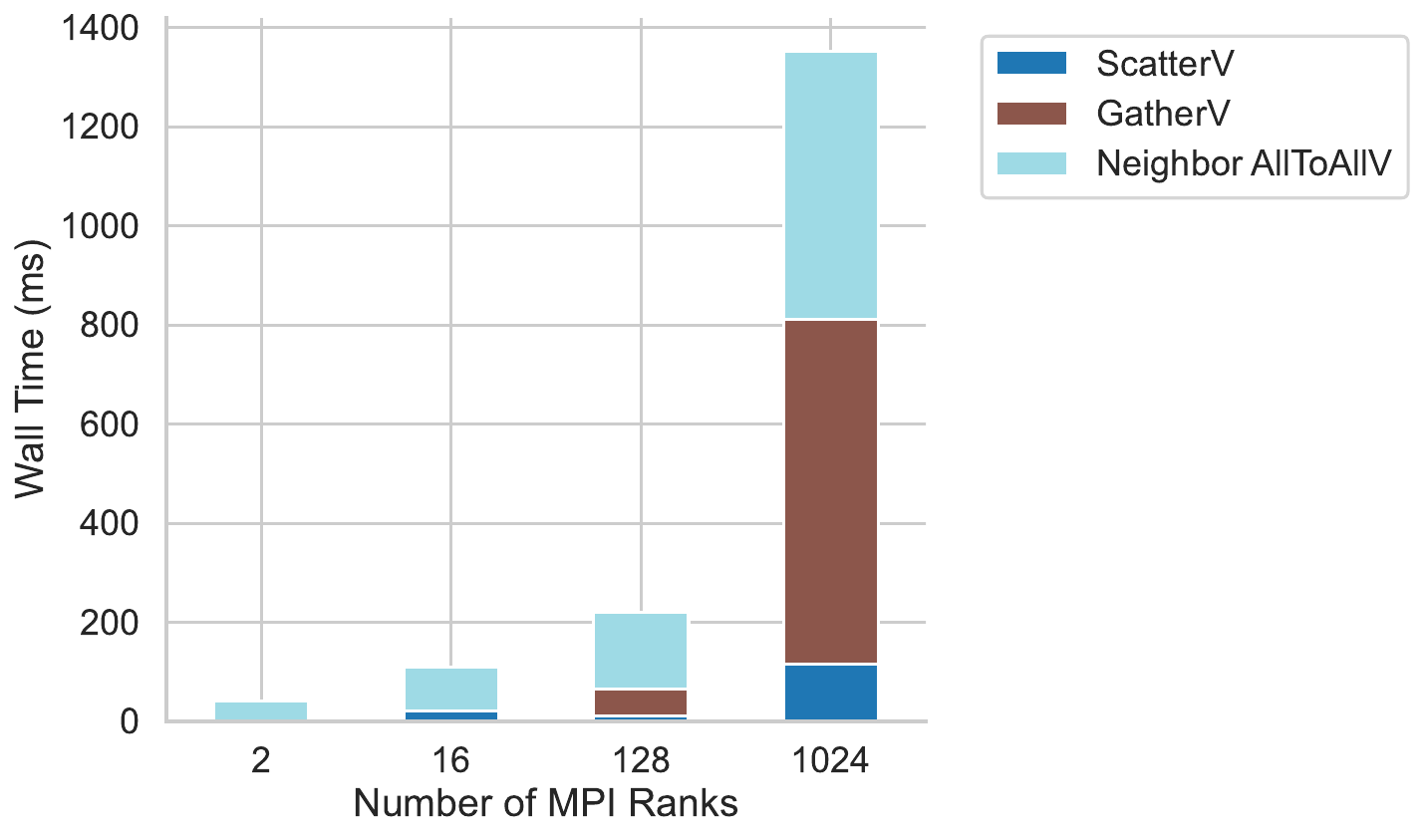}
  \caption{Breakdown of communication time for the same socket-pinned weak-scaling configurations used above. Top: uniform random distribution. Bottom: spherical surface distribution. The stacked bars report mean communication times over five repeated runs and are plotted for illustration; from bottom to top they show \texttt{MPI\_Scatterv}, \texttt{MPI\_Gatherv}, and \texttt{MPI\_Neighbor\_Alltoallv}.}
  \label{fig:comm_breakdown}
\end{figure}

Figure~\ref{fig:comm_trend} partitions the total wall time into communication and computation for the same socket-pinned weak-scaling configuration considered above for scaling up to $3.2e10$ particles. Communication time is measured as wall-clock time spent in the main runtime collectives, namely \texttt{MPI\_Neighbor\_Alltoallv}, \texttt{MPI\_Gatherv}, and \texttt{MPI\_Scatterv} in the runtime phase of our algorithm described in Section \ref{sec:fmm:sub:algorithm}; computation time is the corresponding wall-clock time spent in the \fmm kernel phases. The stacked bars in Figure~\ref{fig:comm_trend} report mean communication and computation times over the five repeated runs, and Figures~\ref{fig:comm_trend} and \ref{fig:comm_breakdown} are included to illustrate how the runtime and communication decompose under this same setup. In both distributions, total wall time is dominated by computation, while communication grows with scale.

This is consistent with the design choice of centralizing the global tree traversal: although global collectives are introduced, they remain a minority of the total runtime throughout the uniform study and across most of the non-uniform study. The effect is more pronounced for the spherical distribution, where irregular occupancy increases communication costs and makes communication a much larger fraction of the runtime at the largest scales.

These measurements also provide the clearest experimental connection to the complexity model of Section~\ref{sec:analysis}. There, the favorable bound relies on two practical conditions: that neighborhood communication remains sparse, and that the global gather/scatter steps remain in the small-message regime where modern \mpi implementations use efficient collective algorithms. Figure~\ref{fig:comm_trend} is broadly consistent with this picture over the measured range for the uniform distribution. This begins to break down for the non-uniform study at the largest problem size measured, however still doesn't quite reach the linear trend-line. It appears likely that increasing the number of \mpi ranks further would see communication complexity move to a worse regime for non-uniform point distributions under our algorithm due to the imbalance in the global collective calls. However, as we show here, we are still able to compute extremely large problems with practically useful runtimes despite this loss of optimality.

\section{Discussion}\label{sec:discussion}

\subsection{Comparison to $\mathcal{HSDX}$}

The scheme most closely related to ours is the $\mathcal{HSDX}$ approach proposed by Abduljabbar et al. \citep{abduljabbar2017communication}, which also uses MPI neighborhood collectives to encode static communication graphs between adjacent subdomains. However, our design departs from theirs in several important ways, especially in how we handle the exchange of $\mathbsf{u}$ data during the local phase of the tree traversal and our focus on retaining shared memory optimizations developed for the single-node case.

In $\mathcal{HSDX}$, each MPI process $P_\beta$ corresponding to a local root $\beta$ first identifies a halo $h_N$ of up to 26 neighboring processes. To resolve all required far-field \textbf{VLI} interactions, an extended halo $h_F$, consisting of second neighbors of $\beta$, must also be accessed. Communication graphs are constructed such that each process in $h_N$ is responsible for querying a portion of $h_F$. This leads to the creation of multiple neighborhood communicators: one between $P_\beta$ and its direct neighbors ($C_N$), and one between each neighbor in $h_N$ and a subset of $h_F$ (denoted $C_F^i$ for $i \in [1, 26]$).

Their algorithm requires four collective operations to identify and collect ghost data for $P_\beta$:

\begin{enumerate}
  \item An initial \texttt{MPI\_Neighbor\_Alltoall} via each $C_F^i$ to query which \textbf{VLI} interactions exist.
  \item  A corresponding \texttt{MPI\_Neighbor\_Alltoallv} to exchange the associated $\mathbsf{u}$ data.
  \item A \texttt{MPI\_Neighbor\_Alltoallv} via $C_N$ to exchange $\mathbsf{u}$ data for neighbors to $P_\beta$.
  \item A final pair of collective calls via $C_N$ to retrieve \textbf{VLI} data from $h_F$ that has been forwarded to $h_N$ in steps (1) and (2).
\end{enumerate}

Because Abduljabbar et al. do not provide a software artifact, we restrict the comparison here to the communication structure described in the paper rather than to inferred implementation-level overheads. On that basis, our approach differs from $\mathcal{HSDX}$ in several concrete ways:

\begin{enumerate}
  \item We perform the global \fmm pass on a single nominated process, enabling reuse of highly tuned shared-memory kernels with minimal additional implementation effort.
  \item The formulation allows an \mpi process to own multiple local trees, which can in principle be merged and processed jointly in shared memory.
  \item All interaction dependencies are precomputed during setup, so runtime traversal does not require discovery of the communication graph.
  \item We use only a single call to \texttt{MPI\_Neighbor\_Alltoallv} at runtime for local ghost exchange, compared to the four collective phases described for $\mathcal{HSDX}$. This is possible because we precompute the full communication layout and separate the ghost layers for \textbf{ULI} and \textbf{VLI}.
  \item Global communication is reduced to small gather/scatter messages that can be handled by standard \mpi collectives.
\end{enumerate}

\subsection{Limitations}

Our method deliberately targets uniform octrees. This design choice simplifies the construction of static communication graphs and allows for fully precomputed interaction lists and data exchanges. As a result, we avoid runtime filtering, memory allocation, and communication planning. In doing so, we sacrifice adaptability to nonuniform trees in favor of maximal precomputation and memory reuse. While this limits generality in comparison to \citep{lashuk2009massively,abduljabbar2017communication}, it enables the full exploitation of shared-memory hardware on uniform problems – and moderately irregular point distributions are also handled well by shared-memory optimizations for the \textbf{ULI} step – which is typically the most straightforward phase of the \fmm to parallelize.

We also introduce a controlled bottleneck by executing the global portion of the \fmm in a single shared-memory execution context. This is a conscious tradeoff: by centralizing global tree traversal, we preserve the performance benefits of shared-memory kernels, particularly for the \textbf{VLI} step. As shown in the ARCHER2 experiments of Section~\ref{sec:benchmarks}, this choice does not appear to hinder scalability over the ranges studied: we observe useful weak scaling together with useful practical strong-scaling behavior on smaller configurations.

Our approach may lose optimal worst-case complexity with $P$ depending on the point distribution, with a clear trade-off between the number of points handled per-rank (determined by the efficiency of our shared memory implementation) and the total number of ranks. However, the clear strength of our approach is the ease with which it can be implemented in \fmm software already developed and optimized for a the single-node case, using only standard collective operations.

\section{Conclusion}\label{sec:conclusion}

We have presented a communication scheme for distributed \fmms based on standard \mpi neighborhood collectives, uniform trees, and precomputed communication layouts. The central goal of the method is to simplify the distributed-memory extension of existing high-performance \fmm software while preserving the shared-memory kernels that dominate runtime on modern machines.

The ARCHER2 results show that this design delivers weak scaling for uniform and moderately non-uniform point distributions, together with useful practical strong-scaling behavior on smaller configurations. The distributed \fmm runtime remains on the order of tens of seconds even at the largest scales studied. Although favorable runtime communication complexity can still be obtained in the best case, the main contribution of this work is to show that a distributed \fmm can be realized using standard \mpi collectives and precomputed communication layouts while still delivering good large-scale runtime performance.

Future work would involve deploying our software on other leading HPC systems, to identify whether the issues we faced with system noise were unique to ARCHER2's communication subsystem, and identify the with more granularity the reasons for the departure from optimal communication complexity in the case of non-uniform point distributions at large $P$.

\bibliographystyle{SageH}
\bibliography{distributed-fmm}
\end{document}